\documentclass[12pt]{iopart}

\usepackage{iopams}

\eqnobysec
\newcommand{\beq}{\begin{equation}}
\newcommand{\eeq}{\end{equation}}
\newcommand{\bea}{\begin{eqnarray}}
\newcommand{\eea}{\end{eqnarray}}

\newcommand{\rd}{\mathrm{d}}            
\newcommand{\re}{\mathrm{e}}            
\newcommand{\ri}{\mathrm{i}}

\newcommand{\te}[1]{\mbox{\boldmath $ #1 $}}
\newcommand{\tes}[1]{\mbox{\boldmath ${\it #1 }$}}
\newcommand{\bx}{\te{x}}
\newcommand{\by}{\te{y}}
\newcommand{\bz}{\te{z}}
\newcommand{\bn}{\te{n}}
\newcommand{\bbs}{\te{s}}

\newcommand{\bu}{\te{u}}
\newcommand{\bv}{\te{v}}

\newcommand{\bbr}{\te{r}}

\newcommand{\bF}{\te{F}}
\newcommand{\bT}{\te{T}}
\newcommand{\bU}{\te{U}}

\newcommand{\bR}{\te{R}}
\newcommand{\bV}{\te{V}}

\newcommand{\bcalU}{\mbox{\boldmath$\cal U$}}

\newcommand\eg{\textit{e.g.}\ }
\newcommand\ie{\textit{i.e.}\ }

\usepackage{graphicx}
\usepackage{dcolumn}
\usepackage{color}

\begin{document}

\title{Undulatory locomotion of finite filaments: lessons from \emph{C. elegans}}
\author{R.~S.~Berman$^1$, O.~Kenneth$^{1,2}$, J.~Sznitman$^{3,5}$ and A.~M.~Leshansky$^{4,5}$}
\address{$^1$Department of Physics, Technion--IIT, Haifa 32000, Israel \\
$^2$School of Physics and Astronomy, Raymond and Beverly Sackler Faculty of
Exact Sciences, Tel-Aviv University, Tel-Aviv 69978, Israel. \\
$^3$Department of Biomedical Engineering, Technion--IIT, Haifa 32000, Israel \\
$^4$Department of Chemical Engineering, Technion--IIT, Haifa 32000, Israel \\
$^5$Technion Autonomous System Program (TASP), Haifa 32000, Israel}
\ead{lisha@technion.ac.il}

\date{\today}

\begin{abstract}
Undulatory swimming is a widespread propulsion strategy adopted by many small-scale organisms including various single-cell eukaryotes and nematodes. In this work, we report a comprehensive study of undulatory locomotion of a finite filament using (i) approximate resistive force theory (RFT) assuming a local nature of hydrodynamic interaction between the filament and the surrounding viscous liquid, and (ii) particle-based numerical computations taking into account the intra-filament hydrodynamic interaction. Using the ubiquitous model of a propagating sinusoidal waveform, we identify the limit of applicability of the RFT and determine the optimal propulsion gait in terms of (i) swimming distance per period of undulation and (ii) hydrodynamic propulsion efficiency. The occurrence of the optimal swimming gait maximizing hydrodynamic efficiency at finite wavelength in particle-based computations diverges from the prediction of the RFT. To compare the model swimmer powered by sine wave undulations to biological undulatory swimmers, we apply the particle-based approach to study locomotion of the model organism nematode \emph{Caenorhabditis elegans} using the swimming gait extracted from experiments. The analysis reveals that even though the amplitude and the wavenumber of undulations are similar to those determined for the best performing sinusoidal swimmer, \emph{C. elegans} overperforms the latter in terms of both displacement and hydrodynamic efficiency. Further comparison with other undulatory microorganisms reveals that many adopt waveforms with characteristics similar to the optimal model swimmer, yet real swimmers still manage to beat the best performing sine-wave swimmer in terms of distance covered per period. Overall our results underline the importance of further waveform optimization, as periodic undulations adopted by \emph{C. elegans} and other organisms deviate considerably from a simple sine wave.

\end{abstract}

\submitto{\NJP}


\maketitle

\section{Introduction}

In the limit of low Reynolds numbers, defined as $Re=U l \rho/\mu\ll 1$, where $U$ is a characteristic speed, $l$ a characteristic length, and $\rho$ and $\mu$ are, respectively, the fluid's density and dynamic viscosity, the locomotion of microorganisms is governed by small length scales such that linear viscous forces typically dominate over nonlinear inertial forces \cite{lighthill76,brennen1977,lauga2009}. For Newtonian fluids in the absence of inertia, the equations of fluid motion are time-reversible and net forward swimming results from non-reciprocal gait to break symmetry; a property best known as the ``scallop theorem'' \cite{purcell1977}. Among the various strategies nature has opted for, undulatory gaits featuring the propagation of planar traveling waves characterize a wide range of small-scale organisms including single-cell flagellates \cite{brennen1977}, various sperm cells \cite{gray1955,GH55}, as well as multi-cellular organisms such as nematodes \cite{cohen2010,gray1964}.

In particular, due to its easiness of manipulation and convenient size, the well-known roundworm {\it Caenorhabditis elegans} ({\it C. elegans}) has gained considerable attention over the past few years as an attractive living model to study experimentally the coupling between small-scale propulsion and low-Reynolds-number hydrodynamics \cite{cohen2010,celegans10}. One defining feature observed in the locomotion patterns of {\it C. elegans} is the robustness of its swimming gait. While it may opt to modulate its locomotory gait in response to the properties of the physical media in which it is immersed, {\it C. elegans} exhibits limited changes in the overall spatial characteristics of its gait. Namely, amplitude ($b$), wavelength ($\lambda$), and importantly forward speed ($U$), remain nearly constant when swimming in Newtonian fluids over a range of viscosities spanning nearly a hundred-fold \cite{celegans10,sznitman2010}, whereas undulating frequency (${\it\Omega}$) shows a very slow monotonic decay. Similarly, for propulsion through non-Newtonian viscoelastic media, such as in aqueous solutions of gelatin \cite{berri2009} or polysaccharide \cite{fangyen2010}, frequency, wavelength, and amplitude of the flexural wave have been observed to decrease slowly within a limited range upon increasing the concentration of the thickening agent. These experimental observations raise the question as to whether \emph{C. elegans'} choice of a specific spatial gait arises as a well-adapted solution to swimming at low Reynolds number, and more generally if there exits optimal swimming gait for planar undulatory locomotion.

Considerable efforts have been pursued to quantify swimming gaits on the basis of various swimming efficiency definitions. For instance, one classic metric compares power expenditure in swimming over a fixed distance at a fixed velocity to the power required to drag the swimmer at the same velocity by an external force (the hydrodynamic efficiency $\delta$ based on that definition is provided below in Eq.~\ref{eq:effdef}). Lighthill considered  hydrodynamic efficiency for locomotion powered by the passage of periodic waves down the length of an infinitely long flagellum and found that for an optimal flagellar waveform, the angle between the local tangent to the flagellum and the swimming direction should be constant (in absolute value) \cite{lighthill75}. Thus, for planar undulations of infinitely-long swimmers the optimal waveform is non-smooth and adopts a sawtooth form. In contrast, for finite slender swimmers, Pirroneau and Katz \cite{PK74} considered the optimal swimming waveform by applying an approximate resistive force theory (RFT) and arrived at the optimal ratio between the amplitude and the wavelength for the saw-tooth and for small-amplitude sinusoidal waveforms. The optimal sawtooth, sinusoidal, curvature sinusoidal and other waveforms of finite filaments have also been studied numerically  using \eg boundary integral approach \cite{SH72,higdon72,DKB80,JB79}, and variants of slender body theory (SBT) \cite{KR76,johnson80}. Spagnolie and Lauga \cite{SL10} considered the regularization of Lighthill's sawtooth waveform while taking into account the additional costs of bending, sliding of the internal microtubules and internal viscous resistance. Most recently, Koehler \emph{et al.} \cite{KST12} reported a detailed numerical study of undulatory locomotion of finite filament in a range of lengths and actuation parameters using RFT.

Yet, one should consider whether or not microorganisms are indeed concerned about the power expenditure in swimming. Experiments and supporting predictions for flagellated bacteria, such as {\it E. coli}, show that locomotion accounts for only a few percent of their metabolic costs \cite{SL11,chat2006}. Hence, if microorganisms are less concerned about hydrodynamic power expenditure, they may care about getting furthest away over a stroke.
For undulatory locomotion driven by a traveling wave propagating along the filament length, the net distance traveled per period can be taken as an alternative measure of propulsion efficiency. Note, however, that for a more general swimmer, stroke or swimming gait, the distance per stroke may not be an adequate metric for comparison; low-Reynolds-number locomotion is geometric such that the net distance covered per stroke is independent of how fast the stroke is. However, for a given frequency of undulation (\eg, the undulation frequency of \emph{C. elegans} may vary from approximately 2~Hz down to less than 0.3~Hz as the solvent viscosity is increased by 10,000 folds \cite{berri2009,fangyen2010}), the only way to move furthest is through optimizing the waveform.

\begin{figure}[t]
\begin{center}
\includegraphics[scale=1]{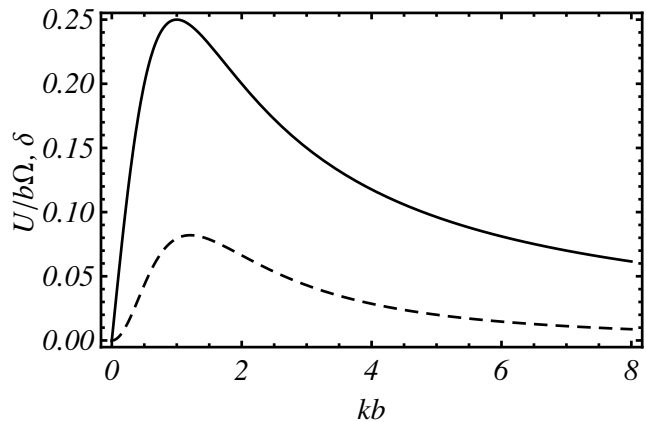} 
\end{center}
\caption{Prediction of the local RFT for an infinitely (exponentially) thin filament with $\xi=2$ propagating traveling sine wave, the solid line stands for the scaled speed of propulsion $U/\it{\Omega} b$ and the dashed line denotes the hydrodynamic efficiency $\delta$.  \label{fig:intro}}
\end{figure}

Before detailing the mathematical models employed here, it is instructive to briefly point out differences in optimal performance pertaining to the two metrics introduced above. For this, let us consider the simplest possible planar sinusoidal waveform. From the point of view of hydrodynamic efficiency, without considering the additional costs associated with bending, internal resistance and others, RFT for undulatory propulsion of an infinite filament (see \cite{GH55} or Sec.~\ref{sec:math} for more details) suggests that there is an optimal product of the amplitude $b$ and the wavenumber $k=2\pi/\lambda$, namely $\kappa=kb\approx1.208$, that maximizes the hydrodynamic efficiency giving $\delta\simeq8.2$~\% (see the dashed line in Fig.~\ref{fig:intro}). Namely, optimized propulsion driven by short small-amplitude waves is equivalent (efficiency-wise) to swimming with long large-amplitude waves, as long as the value of $\kappa=kb$ is maintained at the optimum. For a finite filament of length $l$, however, there is a constraint relating $k$, $b$ and the number of waves $p$ per distance from head-to-tail, such that the increased amplitude would result in a smaller value of $p$ and could lead to considerable pitching and transverse motion, presumably yielding a reduction in hydrodynamic efficiency. Therefore, swimming with many short small-amplitude waves is expected to be the best strategy efficiency-wise for a finite filament. This is in agreement with most recent findings in \cite{KST12}, where RFT was applied to study optimal locomotion of finite filaments for various periodic waveforms. However, RFT does not take into account hydrodynamic intra-filament interaction that could deteriorate hydrodynamic propulsion efficiency when swimming with many short waves.

The situation is different if the optimal displacement per stroke (or the mean propulsion speed) is concerned. The approximate expression for the velocity of an infinite filament propagating a sinusoidal wave based on RFT \cite{GH55} reads (see also Sec.~\ref{sec:rft_infi})
\beq
{U\over c}=-\frac{1}{2} \kappa^2 \frac{\xi-1}{1+\xi\kappa^2/2}\:,\label{eq:speed}
\eeq
where $c=\it{\Omega}/k$ is the wave speed, $f_\perp$ and $f_{||}$ are the normal and longitudinal viscous drag coefficients (\ie per unit length of the filament), respectively, and $\xi=f_\perp/f_{||}$ typically varying between $1$ and $2$ for an incompressible Newtonian liquid. This  approximate solution suggests that there is an optimum at $\kappa=(2/\xi)^{1/2}\approx 1$ that maximizes the scaled velocity $U/(\it{\Omega} b) = -\frac{1}{2} \kappa \frac{\xi-1}{1+\xi\kappa^2/2}$ (see the solid line in Fig.~\ref{fig:intro} for $\xi=2$ corresponding to an exponentially thin filament).

Therefore, increasing the wavelength $\lambda$ and the amplitude of undulations $b$ proportionally to each other at fixed $\it{\Omega}$, while keeping $\kappa$ at the optimum, would yield a faster propulsion. For a finite filament, however, such an upscale of the waveform would result in a smaller value of $p$ leading to considerable pitching motion as $p$ diminishes, presumably hindering propulsion. Since small-amplitude undulations are inefficient speed-wise (the velocity in Eq.~\ref{eq:speed} is quadratic in the amplitude $b$ at $\kappa \ll 1$), there should be an optimal amplitude ($b/l$) and wavelength ($kl$) for a finite-length filament yielding the maximum displacement per period of undulation.

Thus, for a finite undulating filament the two relevant metrics of self-propulsion (\ie the distance covered per stroke and work invested in propulsion) are expected to yield different values of the optimal amplitude and wavelengths. While maximizing the distance covered per stroke determines some particular combination of  $kl$ and $b/l$, power saving strategies require many short small-amplitude waves (at least within the RFT approximation) so that the optimum is expected to be found at the maximum allowable value of $kl$ at the boundary of $(b/l,\:kl)$ domain. An interesting question concerns how far the two optima are separated in the plane of parameters $(b/l,\:kl)$ for finite swimmers and whether undulatory microorganisms including sperm cells, flagellates and nematodes prefer one hydrodynamic efficiency metric over the other. Here, we shall address these points in detail both analytically and numerically, using a combination of the approximate RFT and particle-based computations, where the nonlocal nature of hydrodynamic interaction between different parts of the filament is more rigorously accounted for. As in earlier works \cite{SH72,higdon72,DKB80,JB79}, the present study incorporates numerically the nonlocal intra-filament interactions for finite filaments. However, we detail here a comprehensive parametric study of optimal locomotion (considering either definition) in contrast to previous studies where the major accent was placed on different aspects of undulatory locomotion, including accuracy of slender-body theory against less accurate RFT, filament interaction with a passive head (relevant for sperm cells), and non-sinusoidal undulations.

Due to the complexity of the general problem of what would be the optimal waveform as to swim the furthest over a period of undulation, we restrict our discussion to the simplest possible undulatory gait, namely the planar traveling sinusoidal wave. This waveform has been studied extensively in the past and constitutes a crucial propulsion model in an effort to deepen our general understanding of low-Reynolds-number undulatory locomotion. We compare the performance of our particle-based model to the swimming characteristics of the nematode {\it C. elegans} obtained from experiments \cite{celegans10,sznitman2010} and extend our discussion and results to a wider range of undulatory microorganisms, including nematodes, sperm cells and primitive flagellates.

\section{Mathematical formulation \label{sec:math}}

\subsection{RFT for a finite filament \label{sec:rft_fin}}

The shape of the swimmer at the moment $t$ is given by $s\mapsto \bbr_0(s,t)=\left\{x_0(s,t),y_0(s,t)\right\},\; s_1\leq s\leq s_2$ (see Fig.~\ref{fig:schematic}).
The actual embedding of it in $\mathbb{R}^2$ is given by  $s\mapsto r(s,t)=\left\{x(s,t),y(s,t)\right\}$ where
\[
\bbr(s,t)=\bcalU(t)\cdot\left[\bbr_0(s,t)+\bR(t)\right],\quad \bcalU(t)={\left(\begin{array}{cc} \cos\theta(t) & -\sin\theta(t) \\ \sin\theta(t) & \cos\theta(t) \end{array}\right )}.
\]
The angular velocity of the swimmer is $\te{\omega}=\dot{\theta}\widehat{\te{z}}$, where dot stands for time derivative, and through some abuse of notation one may write $\dot{\bcalU}=\te{\omega}\times{\bcalU}$.
The local swimmer velocity then reads
\[
\bv(s,t)={\rd\bbr\over\rd t}=\te{\omega}\times \bbr+\bcalU\cdot(\dot{\bbr}_0+\dot{\bR})=
\bcalU\cdot\left(\te{\omega}\times\bbr_0+\bV+\bv_0\right)\:.
\]
Here we denote by $\bv_0=\dot{\bbr}_0$ the local deformation velocity and by $\bV=\bcalU^{-1}\cdot{\rd\over \rd t}(\bcalU\cdot\bR)=\te{\omega}\times \bR+\dot{\bR}$ the extra rigid translation experienced by the swimmer both expressed in a frame rotating with it. We shall denote $\gamma=|\bbr_0'|$ where prime stands for ${\partial\over\partial s}$. Then  $\hat{\bbs}={\gamma}^{-1}\bbr_0'$ is the unit tangent to the filament as expressed in  a frame rotating with it. (In the lab frame the unit tangent is $\bcalU \cdot \hat{\bbs}$.)
\begin{figure}[t]
\begin{center}
\includegraphics[scale=0.7]{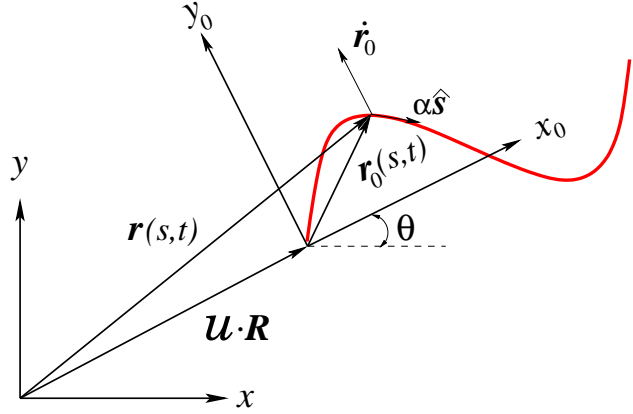} 
\end{center}
\caption{Schematic of the employed coordinate frame (laboratory frame $(x,y)$ and co-moving frame $(x_0,y_0)$) and an undulating filament (red).  \label{fig:schematic}}
\end{figure}
The local velocity $\bv=\dot{\bbr}$ can be written as a sum of parallel and transverse velocities, $\bv=\bv_\|+\bv_\perp$ where
\[
\bv_\|={\dot{\bbr}\cdot\bbr'\over |\bbr'|^2}\:{\bbr'}=\hat{\bbs}\cdot(\bV+\te{\omega}\times \bbr_0+\bv_0)\; \bcalU\cdot\hat{\bbs}
\]
We assume that the local force (per unit length) exerted on the swimmer may be expressed as $\te{f}=f_\perp\bv_\perp+f_\|\bv_\|$ for some
constant $f_\perp,\:f_\|$ and denote $\xi=f_\perp/f_\|$. This allows to express the force (per unit length) as ${1\over f_\|}\te{f}=\xi\bv+(1-\xi)\bv_\|$.
Using the above expressions for $\bv,\bv_\|$ we obtain that the moving frame force $\te{f}_0=\bcalU^{-1}\cdot \te{f}$ is given by
\beq\label{ff}
{1\over f_\|}\:\te{f}_0=
\xi(\bV+\te{\omega}\times \bbr_0+{\bv_0})+(1-\xi)\;\hat{\bbs}\cdot(\bV+\te{\omega}\times \bbr_0+{\bv}_0)\;\hat{\bbs}
\eeq
The zero net force and zero net torque conditions are then
\[
\bF=\int_{s_0}^{s_1}\te{f}_0\; \gamma \rd s=0\:,\qquad \bT=\int_{s_0}^{s_1}\bbr_0\times\te{f}_0\; \gamma \rd s=0\:.
\]
where $\gamma\rd s\equiv\rd \zeta$ is a length element. At each instant $t$ this gives a set of three linear equations for $\bV=(V_x,V_y)$ and $\te{\omega}=\omega \widehat{\te{z}}$. Integration over $t$ then gives $\theta=\int\omega \rd t$ which defines the matrix $\bcalU(t)$. The distance covered by the swimmer is found from $\bcalU\cdot \bR=\int \bcalU\cdot \bV \rd t$.

In our description of the swimmer as $\bbr_0(s,t)$, we parameterized it using a parameter $s$. It was implicitly assumed that each specific value of $s$ corresponds to specific material point of the swimmer. \ie a specific material point at $\bbr_0(s,t_2)$ at time $t_2$ is the same one which was at $\bbr_0(s,t_1)$ at time $t_1$. If this assumption fails, then the calculation described above would fail too. In most biological cases the filament is assumed to be incompressible. This automatically implies that a good parametrization corresponding to actual material points is by its proper length parameter. In such case the correct parametrization should be through the proper length $\zeta=\int\gamma \rd s=\int\left|{\partial \bbr_0\over\partial s}\right| \rd s$  rather then by $s$.
The above formulation would still hold provided we interpret $\bv_0\equiv\dot{\bbr}_0$ as a derivative at constant proper length $\zeta$ rather then at constant $s$,
\beq\label{r0t}
\bv_0=\dot{\bbr}_0=\left({\partial \bbr_0\over \partial t}\right)_\zeta=\left({\partial \bbr_0\over \partial t}\right)_s+\left({\partial \bbr_0\over \partial s}\right)_t \left({\partial s\over \partial t}\right)_\zeta.\label{r00}
\eeq
Actual implementation of this requires calculating $({\partial s/\partial t})_\zeta$ as a function of $(s,\:t)$ for the prescribed undulating filament.

Alternatively, the velocity in (\ref{r00}) can be expressed as
\beq\label{r0t1}
\bv_0={\partial \bbr_0\over\partial t}+\alpha(s,t)\hat{\bbs}
\eeq
for some $\alpha(s,t)$. In other words, the second term on the r.h.s. of Eq.~\ref{r0t} can be interpreted as an extra tangential velocity (shown in Fig.~\ref{fig:schematic}). Demanding incompressibility requires vanishing of the 1D velocity divergence $\nabla_s\cdot \bv_0=\hat{\bbs}\cdot{\partial \bv_0\over\partial s}=0$.
Solving this equation we find $\alpha(s,t)=-\int\hat{\bbs}\cdot{\partial^2 \bbr_0\over\partial s\partial t}\; \rd s+C(t)$ up to some arbitrary function of time $C(t)$.
The integration constant $C(t)$ may be determined by considering the boundary conditions at the swimmer edges. Note that if $s$ is not proportional to the proper length parameter $\zeta$ then incompressibility constraint also implies that its range $s\in[s_0,s_1]$ must be
time-dependent, $s_0=s_0(t), \:s_1=s_1(t)$. The constraint $l=\int_{s_0}^{s_1}\gamma\: \rd s$ does not determine the endpoint $s_0,s_1$ uniquely. Only by specifying an extra condition (\eg requiring $s_0$ or $s_1$ or their average to vanish) does one completely define the swimming mode. The possible arbitrariness of $s_0(t)$ does not matter however, in the special case of our main interest where $\bbr_0(s,t)$ corresponds to a traveling wave
$\bbr_0=\{s,\phi(k s-\it{\Omega} t)\}$. Indeed any  choice of (periodic) $s_0(t)$ may be compensated by redefining the time parameter as $t'=(\it{\Omega} t-k s_0(t))/\it{\Omega}$
(and applying the `gauge' transformation $\bR(t)\rightarrow\bR(t)-\hat{x}s_0(t)$).
Thus in the following we use the simplest choice namely  $s_0(t)=0$. Since the velocity of the endpoint (which is a material point) at $s=s_0$ is $\bv_0={\partial \bbr_0\over\partial t}+{\partial \bbr_0\over\partial s}{\rd s_0\over \rd t}$ we see that the condition $s_0\equiv0$ imply $\alpha|_{s=0}=0$ and hence $\alpha(s,t)=-\int_0^s \hat{\bbs}\cdot{\partial^2 \bbr_0\over\partial s\partial t}\: \rd s$.
For the specific example $\bbr_0(s,t)=\{s,b\sin(ks-\it{\Omega} t)\}$ we obtain
\beq\label{st}
\alpha(s,t)={\it{\Omega}\over k}\left[  \sqrt{1+\kappa^2\cos^2(k s-\it{\Omega} t)} - \sqrt{1+\kappa^2\cos^2(\it{\Omega} t)}\right]\:,
\eeq
where $\kappa=kb$.

The equation  $l=\int_0^{s_1}\gamma\: \rd s$ determining $s_1(t)$ leads in the case of the sine waveform to
\beq
{kl\over\sqrt{1+\kappa^2}}=\mathrm{E}\left(\it{\Omega} t,{\kappa^2\over 1+\kappa^2}\right)- \mathrm{E}\left(\it{\Omega} t-k s_1(t),{\kappa^2\over1+\kappa^2}\right)\:,\label{eq:S1}
\eeq
where $\mathrm{E}(\varphi, m)=\int_0^{\varphi} (1-m\sin^2{\theta})^{1/2}\:\rd\theta$ is the elliptic integral of the second
kind. Only in the special case where the sine wave contains exactly half integer number $p$ of periods, one finds that $s_1(t)=2p\pi/k$ becomes $t$-independent. In this special case one may relate $\kappa$ and $p$ as
\beq
p={kl\over 4\mathrm{E}(-\kappa^2)} \label{eq:meanp}\:.
\eeq
where  $\mathrm{E}(m)\equiv \mathrm{E}(\pi/2,m)$ is a complete elliptic integral.

In the more general case the number of full waves $p(t)=k s_1(t)/(2\pi)$ varies somewhat during a swimming stroke.
The limiting values $(p_{min},\:p_{max})$ of $p(t)$ during a stroke are provided by the solutions of the two equations,
respectively
\[
\mathrm{E}\left(p\pi, -\kappa^2\right)=\frac{kl}{2},\;\;\;\;
\sqrt{1+\kappa^2}\mathrm{E}\left(p\pi, {\kappa^2\over1+\kappa^2}\right)=\frac{kl}{2}\:.
\]
Throughout the paper $p$ will stand for the mean value averaged over a stroke period that may be estimated quite well by  Eq.~\ref{eq:meanp}. The variation of $p$ during a stroke $\Delta p=p_{max}-p_{min}$ can be well approximated by
\beq
\Delta  p\approx {\kappa^2\over 4 \pi (1 + 0.43 \kappa^2)}|\sin(2\pi p)|\:.\label{eq:varp}
\eeq


\noindent The numerical RFT calculations of finite filament locomotion were performed as follows:
\begin{enumerate}

\item First we fixed numerical values for $b,\:k$ and $\xi$ (we fixed $\it{\Omega}=1$, $l=1$ for all calculations).

\item We calculated the expressions for the force and torque densities in the rotated frame
     $\te{f}_0 ,     \: \bbr_0\times \te{f}_0$
by using Eqs.~\ref{ff}, \ref{r0t1}, \ref{st}. This has three independent components corresponding to the force $f_x,\:f_y$ and torque $n_z$ densities.
We expressed them as $A_{ij} q_j+B_i\;,i=1,2,3$ where $\te{q}=(V_x,V_y,\omega)$.

\item We discretized the time range $0\leq t\leq T=2\pi/\it{\Omega}$ into $N=100$ steps $t_i$. (Few calculations were done with higher $N$ up to 300.)

\item For each $t_i$ we first calculated $s_1(t_i)$ by solving Eq.~\ref{eq:S1} numerically.

We then calculated numerically the integrals ${\cal A}_{ij}=\int_0^{s_1}A_{ij} \gamma\rd s, {\cal B}_i=\int_0^{s_1}A_i \gamma\rd s$ and solved ${\cal A}_{ij}q_j+{\cal B}_i=0$ for the values of instantaneous velocities $\te{q}=(V_x,V_y,\omega)$. We kept a table containing the values $(t_i,V_x(t_i),V_y(t_i),\omega(t_i)),\; i=0,1,2,..N$

\item Interpolating $\omega(t_i)$ we constructed a continuous $\omega(t)$ which was then integrated to define $\theta(t)$ and hence the matrix $\bcalU(t)$.

\item Noting the relation $\bcalU\cdot \bR={\rd\over \rd t}(\bcalU\cdot \bV)$ we constructed the `rotated frame center of mass velocity' $\bU_{cm}(t_i)=\bcalU\cdot \bV(t_i)$. We then interpolated it to a continuous $\bU_{cm}(t)$ and integrated over $t$ to obtain the trajectory of the swimmer over a cycle. The distance covered per stroke is then $D=\left|\int_0^T \bU_{cm}(t)\:\rd t\right|$.

\end{enumerate}

\subsection{RFT for an infinite filament \label{sec:rft_infi}}

The analysis based on the local RFT for an infinite undulating filament can be found elsewhere (\eg \cite{GH55,lighthill75}), however we choose to present our short derivation offering a short route to the closed-form expressions for the propulsion velocity and the power dissipated in swimming.

For an infinitely long incompressible undulatory swimmer it is convenient to use a representation of the local velocity $\bv_0$, which takes full advantage of the symmetry/homogenuity of the problem. This representation will be slightly different than the one used in the previous subsection\footnote{The two differ by gauge and by time parametrization but are equivalent.} which was better suited to use in numerics. Consider a swimmer powered by traveling wave type undulations, $\bbr_0(s,t)=\{s,\phi(k s-\it{\Omega} t)\}$. Incompressibility then requires $\bv_0$ to be a superposition of movement along the filament $-\alpha(t)\hat{\bbs}$ and rigid motion (in general a plane translation and rotation).
Taking advantage of the geometric nature of low Reynolds swimming one may assume $\alpha$ to be time independent. In the case of the traveling sine wave we take $\alpha$ to be the velocity required to travel along a period of $\bbr_0(s)$ over time $T={2\pi\over\it{\Omega}}$ i.e.
\beq\label{alpha}
\alpha={\it{\Omega}\over2\pi}\int_0^\lambda \gamma \rd s={2c\over\pi}\; \mathrm{E}{(-\kappa^2)}\:,
\eeq
where $\lambda={2\pi\over k}$ and $c=\it{\Omega\over k}$ is the phase velocity. Averaging over the trajectory of a material point we have $\langle-\alpha\hat{\bbs}\rangle=-c \hat{\bx}$, \ie the phase speed. Thus $\bv_0=-\alpha\hat{\bbs}+c \hat{\bx}$ will describe the local velocity due solely to the (incompressible) periodic deformation. For small amplitudes this is just $\bv_0=\{0,-b\it{\Omega}\cos(ks-\it{\Omega} t)\}+\mathcal{O}(b^3)$.

If we want to describe a swimmer of finite length then we should also specify the parameter range $s\in[s_0,s_1]$.
The fact that the longitudinal velocity $(\bv_0)_x$ is not exactly zero implies that the location of the edges will contain (small) time dependence $s_0=s_0(t),s_1=s_1(t)$. Since the endpoint are material points, $s_0(t)$ and $s_1(t)$ must be solutions of ${\rd s\over \rd t}=(\bv_0)_x=-\alpha \hat{\bbs}\cdot\hat{\bx}+c$.
This leads to the rather complicated relation ($t_0^{(i)}$ being the integration constants)
$$\mathrm{E}\left(ks_i(t)-\it{\Omega} t,{\kappa^2\over 1+\kappa^2}\right)={\alpha k\over\sqrt{1+\kappa^2}}(t_0^{(i)}-t),\;\;i=0,1$$
This deviates from the $s_0=0$ and Eq.~\ref{eq:S1} for $s_1(t)$  used in the previous section due to the use of different time parametrization.

Now consider a very long incompressible swimmer described by $\bbr_0(s,t)=\{s,b\sin(ks-\it{\Omega} t)\}$.
The small oscillations of the endpoints,  $s_{0,1}(t)= \mathrm{Const}-{b^2 k\over8}\sin(2\it{\Omega} t)+\mathcal{O}(b^4)$, are completely negligible compared to $l$. Thus, in the limit of a long swimmer one may take $s_0,s_1$ as constants and even assume $[s_0,s_1]$ to contain exactly a large integer number $p$ of wavelengths. This assumption considerably simplifies the subsequent calculations.

The local velocity due to deformations is $\bv_0=-\alpha\hat{\bbs}+c\hat{\bx}$  with $\alpha$ given in Eq.~\ref{alpha}. The total local velocity would include also a rigid motion which for an infinitely long swimmer can only be a longitudinal motion along x-axis, as transverse translation and rotation are zero from symmetry. The total local velocity is then   $\bv=\bv_0+U\hat{\bx}=-\alpha\hat{\bbs}+\left(c+U\right)\hat{\bx}$ with its longitudinal and transverse components being $v_\| =-\alpha+(\hat{\bx}\cdot\hat{\bbs})\left(c+U\right)$ and $\bv_\perp=(\hat{\bx}-(\hat{\bx}\cdot\hat{\bbs})\hat{\bbs})\left(c+U\right)$, respectively. The corresponding local force on the swimmer is $\te{f}\propto \bv_\|+\xi \bv_\perp$. The transverse component of the force, $F_y$ as well as the torque $N_z$, vanish by symmetry.
The longitudinal force may be expressed in terms of $\gamma={\rd\zeta\over \rd s}=(\hat{\bbs}\cdot\hat{\bx})^{-1}=\sqrt{1+\kappa^2\cos^2(k s-\it{\Omega} t)}$ as
\bea
F_x=\int f_x \rd\zeta&=&\int \left[ \xi \left(c+U\right)-\alpha\gamma^{-1}+\left(c+U\right)(1-\xi)\gamma^{-2}\right] \rd\zeta= \nonumber \\
&& \int \left[ \xi \left(c+U\right)\gamma-\alpha+\left(c+U\right)(1-\xi)\gamma^{-1}\right]\:\rd s
\eea
Requiring $F_x=0$ determines $U$. Since the integrand is periodic of period $\lambda=2\pi/k$ and since the integration range is assumed to be much larger than the wavelength, $s_1-s_0\gg\lambda$, one may just integrate over one period $\int_0^\lambda (\ldots)\rd s$. Using the identities in Appendix A the swimming velocity is then found to be
\beq
\frac{U}{c}=- \frac{(\xi-1)(\mathrm{E}'-\mathrm{K}')}{\mathrm{K}'+\xi (\mathrm{E}'-\mathrm{K}')}\:,\label{eq:rft2}
\eeq
Here  again $\kappa=kb, c=\it{\Omega}/k,\xi=f_\perp/f_{||}$ and $\mathrm{E}'\equiv\mathrm{E}{(-\kappa^2)}$, $\mathrm{K}'\equiv\mathrm{K}{(-\kappa^2\
)}$ where $\mathrm{K}(m)=\int_0^{\pi/2} (1-m \sin^2{\theta})^{-1/2}\:\rd\theta$ is the complete elliptic integral of the first kind. The minus sign indicates that the filament is propelled in the direction opposite to that of the wave propagation.

Note that the RFT result (\ref{eq:rft2}) (as well as the approximate result in Eq.~\ref{eq:speed}) is expected to be strictly valid for undulations satisfying $\varepsilon a\ll 1$, where $\varepsilon=|\partial \hat{\bbs}/\partial \zeta|=|(\gamma^{-1} \partial/\partial s)^2\: \bbr_0|$ is the local curvature of the filament centerline and $a$ is the filament radius. It can be readily shown that $\varepsilon \le b k^2$, and therefore, we expect for the local RFT to hold as long as $a b k^2 \ll 1$ or in a scaled form $\kappa (kl) \ll \epsilon^{-1}$, where $\epsilon=2a/l \ll 1$ is the filament aspect ratio. The latter requirement is less severe than $\kappa \ll 1$ and therefore Eq.~\ref{eq:rft2} is expected to hold for large $\kappa$ as well. Note furthermore that the expression in Eq.~\ref{eq:rft2} is a sole function of $(\mathrm{E}'-\mathrm{K}')/\mathrm{K}'$ which asymptotes to $\sim \kappa^2/2$ at $\kappa \ll 1$. Using this in Eq.~\ref{eq:rft2} yields an asymptotic result which is identical to the approximate solution (\ref{eq:speed}). We also note that the expressions (\ref{eq:speed}) and (\ref{eq:rft2}) also have an identical finite limit for $\kappa\rightarrow\infty$ and therefore, the two expressions provide quite close estimates of the propulsion speed (they differ by at most $\sim$7\% for any $\kappa$ and $\xi$).

It is  interesting to note that a result identical to Eq.~\ref{eq:rft2} may be obtained for an infinitely long sine wave \emph{compressible} swimmer, \ie the swimmer defined by $\bbr=\{s+Ut,\: b\sin(k s- \it{\Omega} t)\}$ and $\bv=\left({\rd\bbr\over \rd t}\right)_s$. This is due to exact cancelations in the integral for the force $F_x$. In the case of a finite length swimmer one must also consider the integrals for the transverse force, $F_y$ and the torque, $T_z$, which usually do not possess similar cancelations. The result for compressible/incompressible case, therefore coincide only in the limit of infinitely long swimmer. To see how the cancelation works for $F_x$ note that the local deformation velocities in the two problems differ in a term of the form $\delta\bv_0=\varphi(ks-\it{\Omega} t)\hat{s}$
for some scalar function $\varphi(ks-\it{\Omega} t)$ whose time average is zero. The extra contribution to the force will be of the same type (up to a multiplicative constant $f_\|$) and therefore $\delta F_x/f_\|=\int\varphi(\hat{s}\cdot\hat{x})\gamma\: \rd s=\int\varphi\: \rd s=0$.

Note that even though Eqs.~\ref{eq:speed},\ref{eq:rft2} were derived for an infinite filament where transverse displacements and pitching cancel out due to symmetry, it can be also applied for approximate modeling of finite-length filament propulsion where transverse displacements and turning are disallowed.

\subsection{Power and hydrodynamic efficiency \label{sec:power}}
The power required for our slender swimmer to maintain its movement is just the dissipation rate $P=\int \te{f}\cdot\bv\: \rd\zeta$.
The total work in a single stroke is $W=\int_0^T P \rd t$. This work depends on the specific time parametrization of the stroke.
It is well known that the optimal (power-wise) time parametrization is the one which makes $P(t)$ time independent, \ie $P(\tau(t))=\mathcal{P}=\mathrm{Const}$. Using the optimal time-parametrization (specifically $\tau(t)=T{\int_0^t\sqrt{P} \rd t'\over\int_0^T\sqrt{P} \rd t'}$) one finds
the optimal work to be\footnote{The fact that $\mathcal{W}\leq W$ may easily be deduced by applying Cauchy-Schwartz inequality. Since the original $t$-parametrization was arbitrary, this proves that $\tau(t)$-parametrization is indeed superior to any other.}
\beq
\mathcal{W}=\int_0^T \mathcal{P} \rd\tau={1\over T}\left(\int_0^T\sqrt{P} \rd t\right)^2\:.\label{eq:work}
\eeq
Our numerical scheme thus allows a simple calculation of $\mathcal{W}$ by integrating $\int_0^T \rd t\sqrt{\int_{s_0}^{s_1} \te{f}\cdot\bv\: \gamma \rd s}$ and squaring it.

For an infinitely long filament we have
$$\te{f}\cdot\bv=f_\perp v_\perp^2+f_\|v_\|^2=f_\perp(U+c)^2(1-\gamma^{-2})+f_\|(\alpha-(U+c)/\gamma)^2$$
Integrating over $s$ and using the identities in Appendix A and Eqs.~\ref{alpha},\ref{eq:rft2} we find:
\beq
P={f_\|c^2 l}\left({4\mathrm{E}'\over\pi^2}-{1\over\xi \mathrm{E}'+(1-\xi)\mathrm{K}'}\right) \mathrm{E}'\:.\label{eq:dissp}
\eeq
(Note that $s_1-s_0={c\over\alpha}l$.) Since the result does not depend on $t$ it is clear that $\mathcal{P}=P$ and the total work per stroke is just $\mathcal{W}=P T$. At $\kappa \ll 1$ the expression in the brackets of (\ref{eq:dissp}) asymptotes to $\frac{\xi  \kappa^2}{\pi}+\mathcal{O}(\kappa^4)$ and $\mathrm{E}'\simeq\frac{\pi}{2}+\mathcal{O}(\kappa^2)$, leading to $\mathcal{P} \approx {1\over2} f_\| \xi c^2\kappa^2l={1\over2} f_\| \xi (\it{\Omega} b)^2 l$.

It is instructive to look at the hydrodynamic \emph{swimming efficiency} $\delta$ that measures the energy dissipated in swimming a fixed distance at a fixed speed as
\beq
\delta=\frac{f_{||} l\: D^2}{T W}.\label{eq:effdef}
\eeq
It resembles the standard Lighthill's propulsion efficiency comparing the power invested in swimming and dragging of inactive filament over distance $D$ with mean velocity $D/T$ \cite{lighthill75}. It is readily seen from Eqs.~\ref{eq:rft2}) and \ref{eq:dissp} that for infinitely long filament the net work and distance per period can be expressed in terms of dimensionless quantities $\widetilde{P}$, $\widetilde{U}$ as $W=\mathcal{W}=\int_0^T P \rd \tau = f_{||} c^2 l T \: \widetilde{P}(\kappa,\xi)$ and $D=U T=c T\; \widetilde{U}(\kappa,\xi)$, yielding
\beq
\delta=\frac{\widetilde{U}^2}{\widetilde{P}}\:.\label{eq:eff0}
\eeq
However, for the finite length filament the hydrodynamic efficiency in (\ref{eq:effdef}) is expressed as
\beq
\delta=\frac{(kl)^2}{(2\pi)^2} \frac{\widetilde{D}^2}{\widetilde{\mathcal{W}}}\:,\label{eq:eff1}
\eeq
where $\widetilde{D}=D/l$ and $\widetilde{\mathcal{W}}=\mathcal{W}/f_{||} c^2 l T$ are, respectively, the dimensionless distance and work per period (corresponding to the optimal time parametrization). Note that in the framework of RFT for either finite or infinite filament both propulsion characteristics,  $D/l$ and $\delta$, do not depend explicitly on $f_{||}$, but are only functions of the ratio $\xi=f_\perp/f_{||}$.

\subsection{Particle-based computations}

The posed problem can be solved using a more accurate (than the local RFT) slender body approximation \cite{cox70,lighthill96}, or numerically, for example, using boundary integral formulation \cite{Pozrik92, Cortez01}. We, however, adopt a different approach and solve the problem using particle-based approach. This technique is based on multipole expansion of the Lamb's  spherical harmonic solution of the Stokes equations (\eg \cite{filippov00}). The filament is constructed from $N$ nearly touching rigid spheres, the so-called ``shish-kebab" model (see Fig.~\ref{fig:schematic}), of radius $a$. The no-slip condition at the surface of all spheres is enforced rigorously via the use of direct transformation between solid spherical harmonics centered at origins of different spheres (see Appendix B). The method yields a system of $\mathcal{O}(N L^2)$ linear equations for the expansion coefficients where the accuracy of calculations is controlled by the number of spherical harmonics (\ie the truncation level), $L$,  retained in the series. This particle-based approach was applied in \cite{LK08,RL08} for modeling Purcell's toroidal swimmer and in \cite{leshansky09} for modeling a propulsion of rotating helical flagellum through a fluid-filled random array of stationary spherical obstacles.
The validity and accuracy of the multipole expansion algorithm was previously tested in \cite{filippov00} against (i) the exact solution (in bi-spherical coordinates) for the flow past two close spheres and against (ii) a boundary element method numerical solution for the translation and rotation of straight chains of spheres (made of $N=2$ to $30$ spheres). For most of our calculations the truncation level $L=2$ yielded quite accurate results as the relative error between the results corresponding to $L=2$ and $L=3$ was less than 5\% even for large-amplitude undulations. A similar approach for particle-based simulations of micro-swimmers based on the extension of Stokesian dynamics was recently proposed in \cite{sd11}. An alternative particle-based approach based on the force-coupling method was applied to  construct a mechanical worm propelled through arrays of micro-pillars \cite{MKZS12}.

The swimming filament is described by $\bbr_0(s,t)=\{s, b\sin{(k s-\it{\Omega} t)+Y(t)}\}$, $s_0\leq s\leq s_1$, where the positions of the endpoints $s_0(t)$ and $s_1(t)$ (\ie centers of the $1$st and $N$th sphere) and the time-periodic function $Y(t)$ are determined from the requirement that the local velocity $\bv_0$ corresponds to a pure deformation \ie the origin of the laboratory coordinate frame is instantaneously fixed with the GC of the filament, $\int_{s_0}^{s_1} \bbr_0 \gamma \rd s =\te{0}$ and the condition $\int_{s_0}^{s_1}  \gamma \rd s =l$ is imposed. Here $\gamma=|\partial \bbr_0/\partial s|$ and $\gamma \rd s$ is a length element of the filament centerline. At each instant, $N$ spheres are positioned equidistant along this centerline in $xy$-plane. The distance between centers of neighboring spheres set equal to $d=2.02 a$ (see the illustration in Fig.~\ref{fig:schematic}). The net filament length is thus fixed as $l=(N-1)\:d+2a$.

The translation velocity ${\bv_0}_i$ of the $i$th sphere due to the centerline deformation is calculated numerically at each time step using a backward difference scheme. ${\bv_0}_i$ consists of transverse undulations $(\partial \bbr_0/\partial t)_i$ plus a tangential velocity $\alpha_i \hat{\bbs}$, as the spheres are re-distributed along the filament due to the incompressibility constraint. The total velocity $\bV_i$ of the $i$th sphere is obtained by adding to ${\bv_0}_i$ an unknown propulsion speed, $\bV$, and rotation with respect to the GC, $\tes{\omega}$. The rotation rate of $i$th sphere composing a filament with respect to its center can be written as $\tes{\omega}_i=\frac{1}{\gamma} (\hat{\bbs}\times\partial \bv_0/\partial s)_i+\tes{\omega}$, where the first term corresponds to the rotation due to local bending and the second term to rigid rotation of the whole filament.

The translation velocity, $\bV=\{V_x(t),\:V_y(t)\}$ and rotation $\tes{\omega}=\omega(t)\hat{\bz}$ are determined from the requirement of force- and torque-free propulsion, \ie $\bF=\sum_i \bF_i=0$, and $\bT=\sum_i \left({\bT}_i+\bR_i\times\bF_i\right)=0$, whereas $\bF=\{F_x,\:F_y \}$ and $\bT=T_z \hat{\bz}$. Here $\bF_i=\int_{\partial S_i} \te{\sigma\cdot}\bn\:\rd S$ is the hydrodynamic force and $\bT_i=\int_{\partial S_i} \te{r}_i\times (\te{\sigma\cdot}\bn)\:\rd S$ is the hydrodynamic torque exerted on $i$th sphere composing the filament. The rate-of-work expended in propulsion of an undulating filament can then be found as
\beq
P=\sum \limits_{i=1}^{N} (-\bV_i\cdot\bF_i-\tes{\omega}_i\cdot \bT_i)\:, \label{eq:power}
\eeq

After calculating the translation and rotation velocities, $V_x(t)$, $V_y(t)$ and  $\omega(t)$, respectively, over a period $2\pi/\it{\Omega}$, we integrated the interpolated velocities over time to compute the trajectory of the filament in $xy$-plane $\bR(t)$:
\beq
\theta(t)=\int_0^t \omega\: \rd t\:,\qquad \bU=\bcalU(\theta)\cdot \bV\:, \quad \left|\bR(t)\right|=\left|\int_0^t \bU(\tau)\: \rd \tau\right|\:,
\eeq
where $\bcalU(\theta)$ is the rotation matrix associated with $\theta(t)$.

We calculated the plane motion and net displacement of the GC of a filament composed of 30-60 spheres, using 100 time steps per period of undulation. We also performed a simplified `1D' calculation in which no pitching or transverse motion was allowed (i.e. $U_y=\omega=0$ was enforced) while $U_x$ was calculated by requiring only $F_x=0$. Since such `1D' calculation was found to be less sensitive to numeric accuracy than the full plane motion, it was sufficient in this case to use only 32 time steps per period of undulation. Note that undulations for which $kl$ is fixed in time were considered. This implies that the number of full waves, $p$, may slightly vary during the stroke period due to the constant length requirement as discussed in Sec.~\ref{sec:math}. This variance, $\Delta p$, can be important at large values of $b/l$, \eg at $b/l \gtrsim 0.8$, the variation can be significant and up to 30\% of the mean value of $p$. However, at $b/l \sim1$ one cannot consider $b$ as an amplitude of the undulation and the swimming gait no longer resembles traveling wave. For biologically relevant swimming gaits with $b/l\lesssim 0.2$ (see Fig.~\ref{fig:organisms}\emph{a}), $\Delta p/p$ remains below 5\% . Recall that the values of $p$ reported in the results correspond to the mean number of waves averaged over a period of undulation.
\begin{figure}[t]
\begin{center}
\includegraphics[scale=0.75]{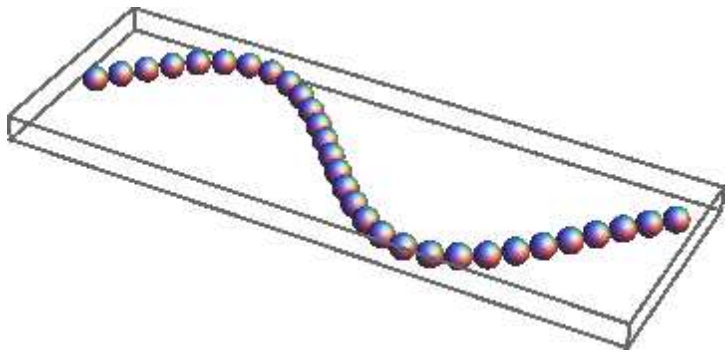}
\end{center}
\caption{Snapshot of the filament built from $N=30$ spheres of radius $\mathtt{a}$ propagating a sine wave with an amplitude $b/l\simeq 0.12$ and $kl\simeq 9.25$ and $p\simeq1.2$ corresponding to the most efficient swimming gait (see Sec.~\ref{sec:res}).   \label{fig:schematic}}
\end{figure}

\section{Results and Discussion\label{sec:res}}

\subsection{Finite sine swimmers: RFT, small-$\kappa$ asymptotic analysis \label{sec:smallkappa}}

The RFT of plane locomotion of the finite sine swimmer is completely determined by the equations of Sec.~\ref{sec:rft_fin}. Finding an analytic solution to these equations is clearly not feasible. However, expanding all variables in a small $\kappa$ Taylor series, it is possible to solve analytically for the leading Taylor coefficients. The resulting approximate solution is expected to be correct up to a relative error of $\mathcal{O}(\kappa^2)$. Comparison with the numerical RFT results shows that it conforms with these formulae for $\kappa \lesssim 0.4$ (see Figs.~\ref{fig:mean_p}\emph{a}--\emph{c}). In particular one finds
\bea
D&\approx& kb^2(\xi-1)\pi\left(1-{4\over k^2 l^2}(2+\cos(kl))+{24\over k^3 l^3}\sin(kl)+ \right.   \nonumber\\
&& \left. {24\over k^4 l^4}(\cos(kl)-1)\right)\:,  \label{eq:Das} \\
\theta_{max} &\approx& -{24 b\over k^2 l^3}\left(kl\cos(kl/2)-2\sin(kl/2)\right)\:, \label{eq:Thetas} \\
\mathcal{W} &\approx& {4f_\|\xi\it{\Omega} b^2\over\pi k^2 l}(k^2 l^2-4(1-\cos(kl))+kl\sin(kl))\mathrm{E}^2(\chi)\:, \label{eq:Was}
\eea
where
\[
\chi={48+8k^2 l^2+16(k^2 l^2-3)\cos(kl)+2kl(k^2 l^2-24)\sin(kl) \over k^2 l^2(k^2 l^2-4(1-\cos(kl))+kl\sin(kl))}\:,
\]
and as before $\mathrm{E}$ stands for the complete elliptic integral. Here $\theta_{max}$ corresponds to the maximum pitching angle, \ie twice the angle of maximum tilt of the rotated frame $\bbr_0(s,t)$ with respect to the direction of propulsion. The asymptotic expression for $\delta$ is too lengthy to give any useful insight. However, it may be of interest to consider the limit of high values of the parameter $p=kl/{2\pi}+\mathcal{O}(\kappa^2 (kl))$, \ie whereas the swimming gait involves multiple waves propagating along the filament.  In this limit the relation simplifies into
\[
\delta\approx \kappa^2{(\xi-1)^2\over 2\xi}\left(1-{127+64\cos(2p\pi)+\cos(4p\pi)\over 64\pi^2 p^2}\right)\:,
\]
having local maxima at half integer values of $p$. Note also that in this case pitching is eliminated since the maximum turning angle $\theta_{max}=-{6\kappa\over\pi^2 p^2}\cos(\pi p)+\mathcal{O}(p^{-3})$ vanishes.
The analogous expansions for $D$, $\mathcal{W}$ at high $p$ read
\bea
D \approx \kappa^2 l{(\xi-1)\over 2p}\left(1-{2+\cos(2\pi p)\over\pi^2 p^2}\right)\:,\nonumber \\
\mathcal{W}\approx \kappa^2{f_\|\xi\Omega l^3\over 4\pi p^2}\left(1-{129+64\cos(2\pi p)-\cos(4\pi p)\over 64\pi^2 p^2}\right)\:. \nonumber
\eea
Note that in the limit $p\rightarrow 1$ the infinite filament small-$\kappa$ results are recovered as expected.

\subsection{Finite sine swimmers: RFT vs. particle-based computations}

The comparison of the local RFT with the results of particle-based computations requires the knowledge of the ratio $\xi=f_\perp/f_\|$. For slender filaments the corresponding viscous drag coefficients (\ie per unit length) are $f_{\perp}=2f_{||}\approx 4\pi\mu E+\mathcal{O}(1)$, where $\mu$ is a dynamic viscosity, $E=(\ln{2/\epsilon})^{-1}$ is a small parameter and $\epsilon=2a/l\simeq 1/N \ll 1$ is the aspect ratio, while $l$ and $2a$ are the length and the typical width of the filament, respectively \cite{KK91}. However, the limiting value of $\xi=2$ is only achieved for extremely slender (exponentially thin) filaments. The classical RFT theories of undulatory locomotion, \eg \cite{GH55} and \cite{lighthill76}, suggest that $\xi$ is slowly varying function of either $\lambda/a$ or $\Lambda/a$, respectively, where $\Lambda$ is the filament length in one full wave, however, these theories assume swimming gaits with many wavelengths per filament length, \ie $\lambda \ll l$, while we look at $p \sim 1$.

\begin{figure}[t]
\begin{center}
\begin{tabular}{cc}
\includegraphics[scale=0.85]{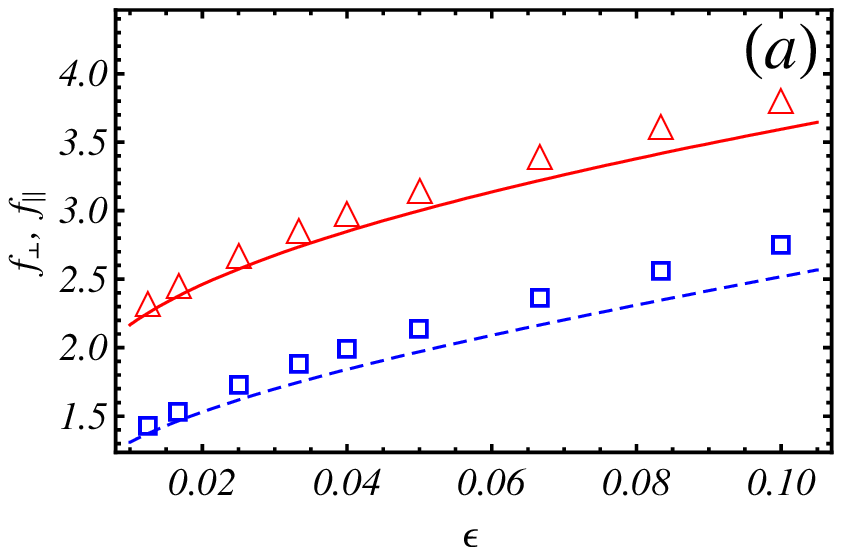} &
\includegraphics[scale=0.85]{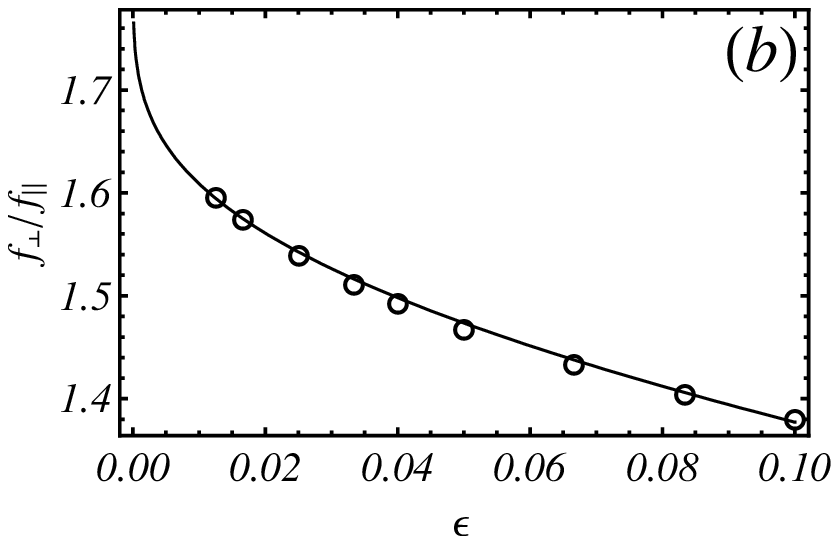}
\end{tabular}
\end{center}
\caption{(\emph{a}) Comparison of the drag coefficients $f_\perp$ ($\triangle$), $f_{||}$ ($\square$) determined via the particle-based computation for a straight chain of length $l$ made of $N=12$ to $80$ nearly touching spheres of radii $a$ (symbols) vs. the predictions of the slender body theory for a prolate spheroid (solid and dashed line, respectively), as a function of the same aspect ratio $\epsilon=2a/l$; (\emph{b}) ratio of the drag coefficients $f_{\perp}/f_{||}$ for a  filament made of spheres as function of the aspect ratio $\epsilon=2a/l\simeq 1/N$ ($\circ$); the continuous line stands for the best fit, $\xi=c_1 (1-c_2 E)/(1+c_2 E)$, with $c_1=1.96$ and $c_2=0.525$ (the slender body theory result for a prolate spheroid \cite{tillet70} corresponds to $c_1=2$ and $c_2=0.5$).  \label{fig:shishkebab}}
\end{figure}

In order to account for the finite thickness of the undulating filament, the value of $\xi$ in  RFT predictions was determined numerically from computing the longitudinal and transverse viscous drag forces for a ``shish-kebab`` straight chain of $N$ nearly touching spheres. The corresponding values of the individual drag coefficients $f_\perp$ and $f_{||}$ as a function of the filament's aspect ratio $\epsilon$ are shown in Fig.~\ref{fig:shishkebab}\emph{a} together with the asymptotic results for a prolate spheroid \cite{tillet70} accurate up to $\mathcal{O}(\epsilon^2\ln{\epsilon})$. Both coefficients $f_\perp$ and $f_{||}$ computed for a ``shish-kebab" filament are slightly larger than the respective coefficients corresponding to a prolate spheroid and the deviation increases with the increase in $\epsilon$ (\ie for less slender filaments). Values of $\xi$ as a function of the rod aspect ratio $\epsilon$ are depicted in Fig.~\ref{fig:shishkebab}\emph{b} ($\circ$). Although there is no analytic or asymptotic theory for such ``shish-kebab" rod, however, the slender body theory solution for a prolate spheroid \cite{tillet70} yields $\xi=2\left(\frac{1-E/2}{1+E/2}\right)+\mathcal{O}(\epsilon^2\ln{\epsilon})$. Approximating our numerical results by the model $\xi \approx c_1 \left(\frac{1-c_2 E}{1+c_2 E}\right)$ (solid line in Fig.~\ref{fig:shishkebab}\emph{b}) suggested by this theory gives $c_1=1.96$ and $c_2=0.525$, which is quite close to the theoretical values ($c_1=2$, $c_2=0.5$) for the prolate spheroid. This indicates that $\xi$ is rather insensitive to the local variation of the filament shape. This is in agreement with \cite{tillet70} where it was shown that for a general slender axisymmetric body the asymptotic result $\xi \approx 2+2(\ln{\epsilon})^{-1}$ is independent of the way in which the cross-sectional radius varies along the length, while the error in $\xi$ estimate due to spatial variance of the local filament shape is of $\mathcal{O}[(\ln{\epsilon})^{-2}]$. The model for $\xi$ indicates that it approaches the limiting value of $2$ logarithmically slow (see Fig.~\ref{fig:shishkebab}\emph{b}) and in the wide range of biologically relevant slenderness $\xi$ is in the range $1.4$--$1.6$. For instance for $N=30$ ($\epsilon\simeq0.033$) we find $\xi \simeq 1.515$.

\begin{figure}[t]
\begin{center}
\begin{tabular}{cc}
\includegraphics[scale=0.85]{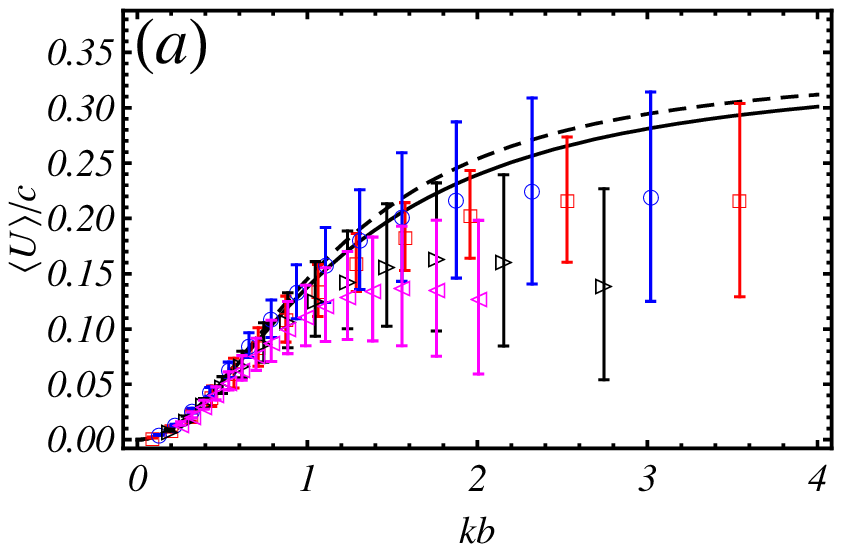} &
\includegraphics[scale=0.85]{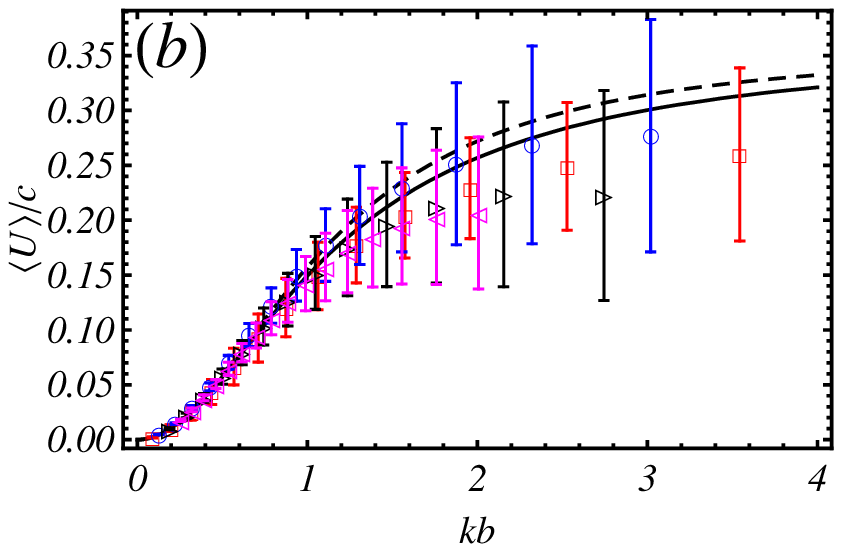}
\end{tabular}
\end{center}
\caption{Scaled averaged (over time of a period) propulsion speed $\langle U\rangle/c$ plotted vs. $\kappa=kb$ for a filament undergoing 1D locomotion opposite to the direction of wave propagation (without pitching and transverse motion).  The symbols stand for the numerical results for $p=0.7$ ($\square$), $p=1$ ($\circ$), $p=1.5$ ($\vartriangleright$) and $p=2$ ($\vartriangleleft$) The bars stand for the root mean square deviation of the swimmer's instantaneous velocity from its mean value during a period of undulation. The solid line corresponds to the prediction of the RFT in Eq.~\ref{eq:rft2}, the dashed line corresponds to the approximate solution (\ref{eq:speed}). (a) filament composed of 30 spheres corresponding to $\xi=1.515$ in RFT expressions; (b) filament composed of 50 spheres, corresponding to $\xi=1.56$ in RFT expressions. \label{fig:Uc_vs_bk}}
\end{figure}

First we test the limits of applicability of the local `1D' RFT analytic result in Eq.~\ref{eq:rft2} towards modeling propulsion of finite-length force-free ($F_x=0$) filament. The scaled propulsion velocity (averaged over a stroke period) for the simplified `1D' model, $\langle U \rangle/c$ together with the root mean square deviation from the mean value (bars), is depicted vs. the amplitude-to-wavelength ratio $\kappa=kb$ in Figs.~\ref{fig:Uc_vs_bk}\emph{a}, \emph{b} for filaments composed of $30$ and $50$ spheres, respectively. As discussed in Sec.~\ref{sec:rft_infi}, the local RFT is expected to be applicable at $\varepsilon a\ll 1$, where $\varepsilon$ is the local curvature of the filament, yielding the condition $\kappa (kl) \ll \epsilon^{-1}$. Our results show that for small values of $\kappa$ all numerical results fall on the theoretical curves in Eq.~\ref{eq:rft2}, while at higher $\kappa$ it may deviate considerably. It is reasonable to expect that upon reducing the width of the filament (or, alternatively, increasing filament's length in particle-based computations) the agreement with the RFT prediction (\ref{eq:rft2}) for the same value of $p$ should improve. Indeed, increasing all lengths by the same factor to preserve $p$, the value of $\kappa (kl)$ remains fixed, while $\epsilon^{-1}$ increases, so the deviation from RFT is expected to kick in at a somewhat higher value of $\kappa$. It can be readily seen in Fig.~\ref{fig:Uc_vs_bk}a, b that the agreement of the numerical results and RFT is closer for a longer filament composed of 50 spheres in comparison with a filament composed of $30$ spheres.

\begin{figure}[tbc]
\begin{center}
\includegraphics[scale=0.85]{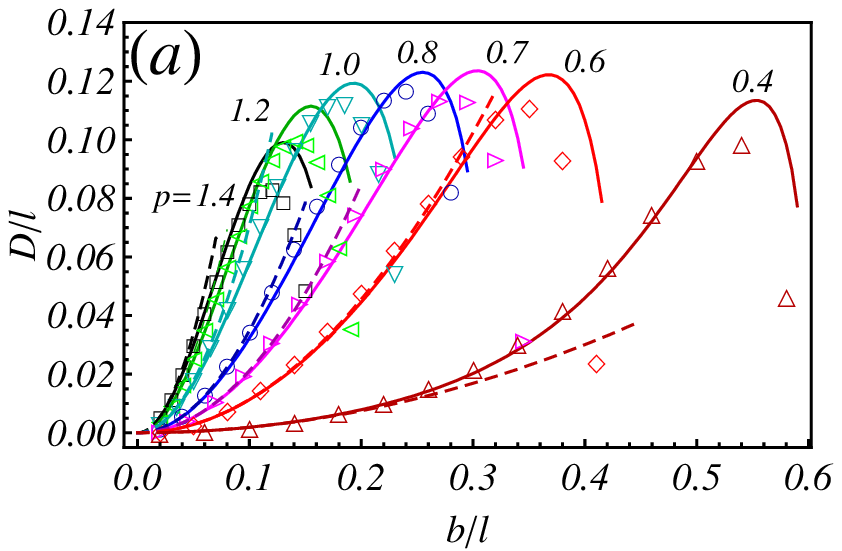}  \\ 
\vskip0.2cm
\includegraphics[scale=0.85]{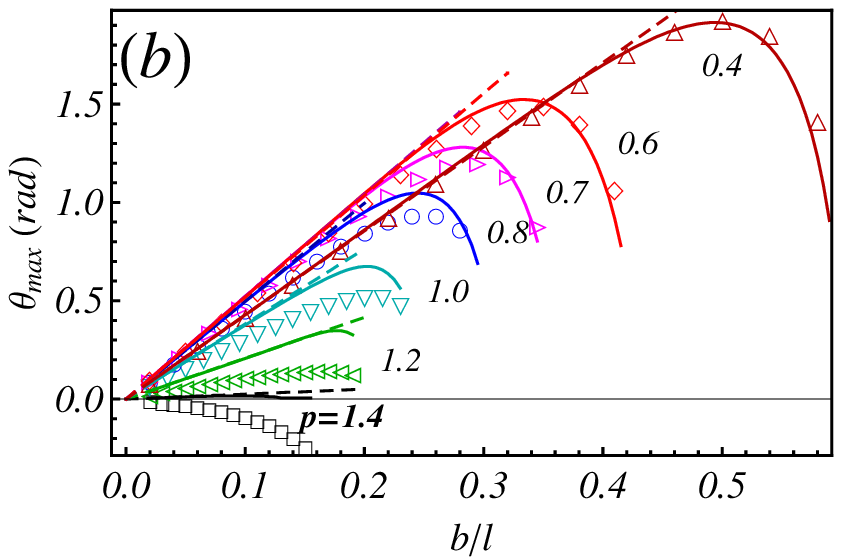} \\ 
\vskip0.2cm
\includegraphics[scale=0.85]{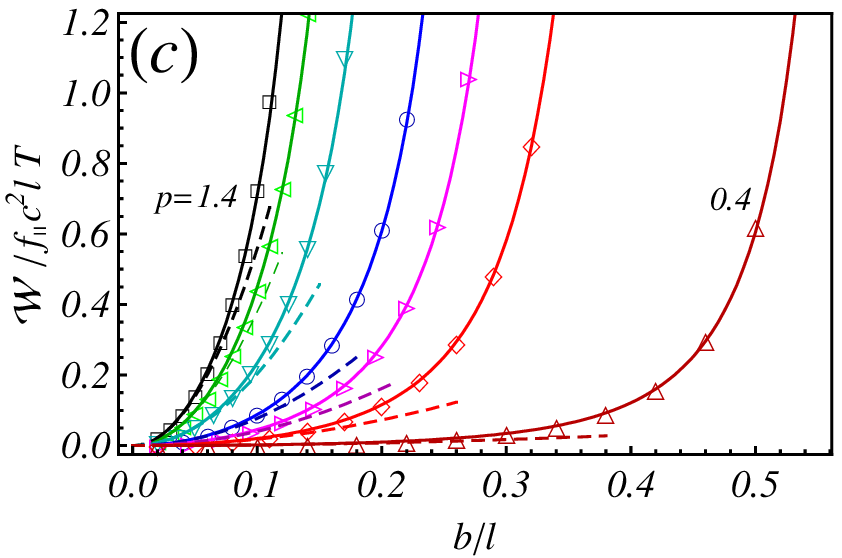} 
\end{center}
\caption{Comparison of the particle-based results vs. the prediction of the RFT for a finite length filament with aspect ratio $\epsilon\simeq 1/30$ upon keeping the mean number of waves fixed (\ie fixed $p$); the corresponding values of mean $p$ are shown. Symbols correspond to the results of particle-based computations: $p=1.4$ ($\square$), $p=1.2$ ($\vartriangleleft$), $p=1$ ($\triangledown$), $p=0.8$ ($\circ$), $p=0.7$ ($\vartriangleright$), $p=0.6$ ($\lozenge$) and $p=0.4$ ($\vartriangle$);  the solid lines correspond to the numerical RFT calculations and dashed lines correspond to the small-$\kappa$ asymptotic RFT predictions in Eqs.~(\ref{eq:Das}--\ref{eq:Was}), both at  $\xi=1.515$. (\emph{a}) The scaled distance per period of undulation, $D/l$ vs. the scaled undulation amplitude $b/l$; (\emph{b}) maximum pitching angle during a cycle, $\theta_{max}$, vs. the scaled amplitude $b/l$; (\emph{c}) optimal work per period of undulation, $\mathcal{W}/f_{||}c^2 l T$, vs. the scaled amplitude $b/l$. \label{fig:mean_p}}
\end{figure}

Next we calculate the net displacement per stoke period for a force- and torque-free plane motion. Fig.~\ref{fig:mean_p}\emph{a} show the scaled distance covered per period, $D/l$, vs. undulation amplitude $b/l$ for different values of mean $p$. Each curve corresponds to a fixed value of mean $p$. For each swimming gait(\ie for each value of mean $p$) there is an optimal amplitude that maximizes the displacement, as was suggested earlier. Note that for the waveforms characterized by $p$ in the range $0.4$--$1.2$ the optimal displacement $D/l$ varies in a relatively narrow range $\sim 1$--$1.2$. The maximum displacement $D/l \approx 0.117$ is achieved at $p\approx 0.8$ and $b/l\approx 0.24$. The agreement between the prediction of the local RFT and the particle-based calculation is quite close in terms of both the optimal amplitude, and displacement, although the RFT  seem to overestimate the displacement at large amplitudes beyond the peak likely due to hydrodynamic self-interaction between the parts of the curved filament which is not taken into account by the RFT. We found that the amplitude at which the deviation between RFT and the particle-based calculations kicks in is well correlated with local curvature of the filament, namely $\kappa (kl)\sim 0.5\epsilon^{-1}$. For large amplitudes such that $\kappa (kl) \gtrsim 0.5\epsilon^{-1}$ the RFT can significantly overestimate the net swimming distance as shown in Fig.~\ref{fig:mean_p}\emph{a}.

Fig.~\ref{fig:mean_p}\emph{b} shows the maximum pitching angle $\theta_{max}$\footnote{This angle can be readily identified in the animations (see supplementary material) as twice the angle between the mean direction of propulsion and the $x$-axis indicating the initial orientation of the rotated frame at $t=0$.} during the stroke period, vs. an amplitude $b/l$ for the same values of $p$ as in Fig.~\ref{fig:mean_p}\emph{a}. It can be seen that for incomplete wave with $p<1$, the pitching angle is considerable and can get to $\theta_{max} \sim 90^\circ$ for $p=0.6$. The agreement between the RFT and the particle-based calculation is reasonable, except at the higher values of $p$ where the curvature is probably too high for the underlying assumptions of RFT to remain accurate. The increase in $p$ yields smaller $\theta_{max}$, as expected, as for many waves we expect no pitching. An interesting observation is that for the finite-length filament the optimal propulsion is associated with substantial pitching during the cycle, as $\theta_{max}\approx 53^\circ$ at $b/l\approx0.24$ for locomotion with $p=0.8$. Note that there are other potential definitions of pitching angle, \eg based on head-to-tail vector which is probably more suitable for image processing of swimming gaits in experiments with undulatory microorganisms, that may produce somewhat different results (at the optimum for $b/l\approx0.24$ and $p=0.8$ it yields pitching angle  $\approx 40^\circ$). Nevertheless, the detailed investigation of alternative definitions and its relevance will be conducted elsewhere.

The scaled net work per period invested in swimming (\ie dissipated by viscosity), $\mathcal{W}/f_{||}c^2 l T$, is plotted vs. the scaled amplitude $b/l$ in Fig.~\ref{fig:mean_p}\emph{c} for the same values of $p$ as in the two previous figures. Note that the parameter $\mathcal{W}/f_{||}c^2 l T$ corresponds to the optimal time parametrization of the stroke defined in (\ref{eq:work}). Obviously, for a fixed amplitude, $b/l$, swimming with a smaller fraction of wave, $p$, is advantageous power-wise, since the relative motion between different parts of the filament diminishes. Note that while the favorable comparison between the RFT and the results of particle-based calculations for $D/l$ and $\theta_{max}$ (in Figs.~\ref{fig:mean_p}\emph{a},\emph{b}) only requires the value of the ratio $\xi=f_{||}/f_{\perp}$, comparison
of the work necessitates knowledge of both $\xi$ and $f_{||}$. We found that the value of $f_{||}$ that fits best the RFT results in the whole range of wavelengths and amplitudes is a sole function of the filament aspect ratio $\epsilon$. This observation deviates from the classical theories \cite{lighthill76,GH55} showing that $f_{||}$ is a slowly (logarithmically) decaying function of the wavelength (either $\lambda/a$ or $\Lambda/a$). However, these theories assume very long filaments, $\lambda \ll l$, while we focus on short filaments, where this assumption may not be valid. For the filament with aspect ratio $\epsilon\simeq 1/30$ we found that $f_{||} \approx 3.30 \mu$ yields an excellent agreement between the prediction of the RFT and particle-based calculations for all values of $p$, while for $\epsilon\simeq1/50$ the corresponding value was $f_{||}\approx 2.97 \mu$. Note that the fitted values of $f_{||}$ are significantly larger than the corresponding values obtained from dragging straight ``shish-kebab" filaments of the same length, \ie $f_{||}\simeq 1.88\mu$ and $f_{||}\simeq 1.61\mu$ for $\epsilon\simeq 1/30$ and $1/50$, respectively. This can be attributed to the increased dissipation resulting from bending a filament made of nearly touching spheres due to the shearing flow in the gaps between neighboring spheres, which does not come into play when dragging a straight ``shish-kebab" filament (which produces $f_{||}$ which is about the same as that for a prolate spheroid, as can be seen in Fig.~\ref{fig:shishkebab}\emph{a}).

It can be readily seen that the small-$\kappa$ asymptotic predictions of $D/l$, $\theta_{max}$ and $\mathcal{W}/f_{||}c^2 l T$ based on RFT in Eqs.~(\ref{eq:Das}--\ref{eq:Was}) shown in Figs.~\ref{fig:mean_p}\emph{a}--\emph{c} as dashed lines match the numerical results at small amplitudes $b/l$ (\ie for $\kappa \lesssim 0.4$) quite well.

The major numerical results gathered in Fig.~\ref{fig:mean_p} can also be re-cast to show net displacement, maximum pitch angle and net work per period as a function of scaled amplitude $b/l$, while keeping the wavelength of the undulation, $kl$, fixed and allowing $p$ to vary. This way we can probe undulations with much larger amplitude since for prescribed $p$ the amplitude grows very slowly with the increase in $kl$ (see Eq.~\ref{eq:meanp} and the dashed red curves corresponding to fixed values of $p$ in Fig.~\ref{fig:contours}). Recall that for $b/l \gtrsim 1$ the swimming gait no longer resembles traveling wave and considering $b$ as an amplitude of the undulation in such case could be misleading.

\begin{figure}[tbc]
\begin{center}
\includegraphics[scale=0.85]{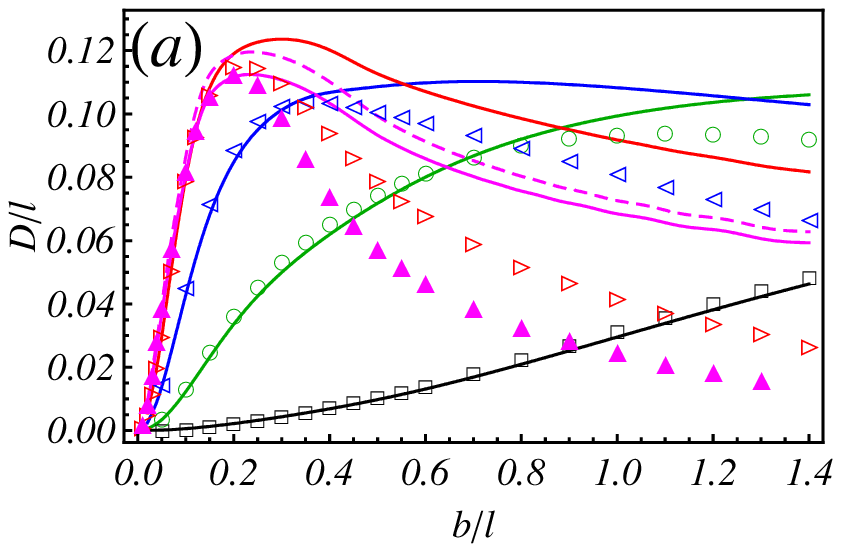} \\ 
\vskip0.2cm
\includegraphics[scale=0.85]{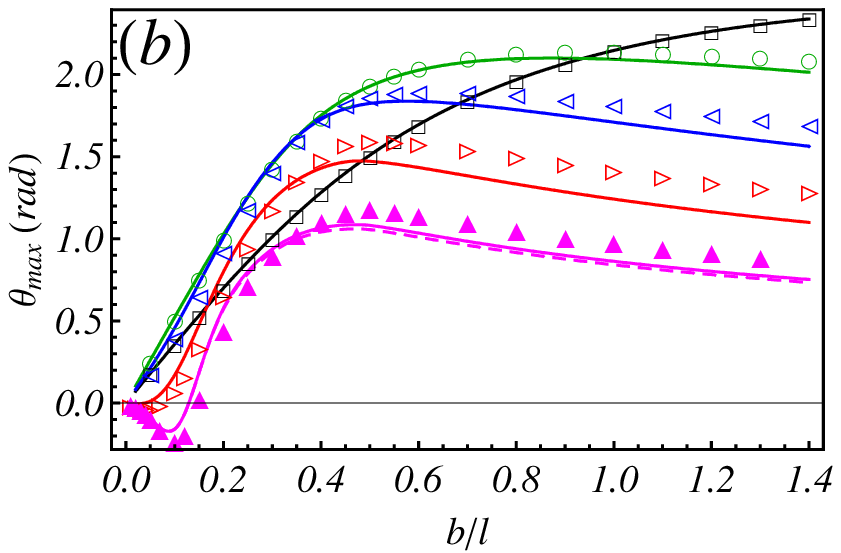} \\ 
\vskip0.2cm
\includegraphics[scale=0.85]{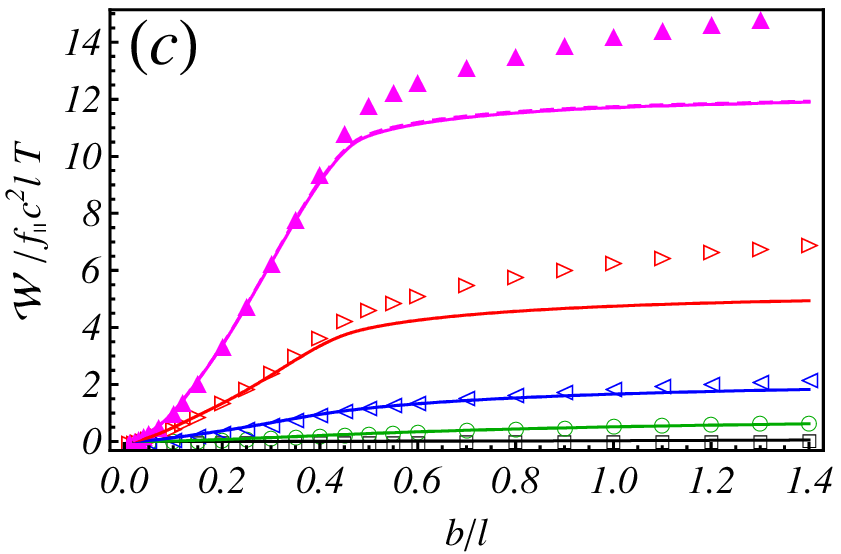} 
\end{center}
\caption{Comparison of the particle-based results (symbols) vs. the predictions of the RFT (lines) for a finite length filament with aspect ratio $\epsilon\simeq 1/30$ upon keeping the fixed value of the wavenumber $kl$: $2$ ($\square$), $4$ ($\circ$), $6$ ($\vartriangleleft$), and $9.25$ ($\vartriangleright$). The upper (filled) triangles ($\blacktriangle$) correspond to $kl=14$ computed for a longer filament with aspect ratio of $\epsilon\simeq 1/50$. The continuous lines stand for the prediction of the RFT with $\xi=1.515$ (corresponding to $\epsilon\simeq 1/30$, solid lines) and $\xi=1.56$ (corresponding to $\epsilon\simeq 1/50$, dashed line). (\emph{a}) the scaled swimming distance per period of undulation, $D/l$ vs. the scaled undulation amplitude, $b/l$; (\emph{b}) maximum pitching angle during a period of undulation, $\theta_{max}$, vs. the scaled amplitude $b/l$; (\emph{c}) optimal work per period, $\mathcal{W}/f_{||}c^2 l T$, vs. the scaled amplitude $b/l$. \label{fig:mean_k}}
\end{figure}

The swimming distance per period of undulation, $D/l$ vs. the scaled amplitude $b/l$ is depicted in Fig.~\ref{fig:mean_k}\emph{a} for several values of $kl$ for a filament with aspect ratio $\epsilon \simeq 1/30$. For $kl=14$ we took a longer filament composed of 50 spheres ($\epsilon\simeq 1/50$) to avoid particle overlap at large amplitudes; the corresponding RFT prediction is not very sensitive to the filament length as it only depends on $\xi$ which is a weak (logarithmic) function of $\epsilon$ (compare the solid and dashed curves corresponding to $kl=14$ in Fig.~\ref{fig:mean_k}\emph{a}). The agreement between the prediction of the RFT and the results of particle-based calculations is very good for moderate values of $kl=2$ ($\square$), $kl=4$ ($\circ$) for amplitudes up to $b/l\sim 1$, while for larger $kl=9.25$ ($\vartriangleright$) the deviation appears at $b/l\sim 0.2$ already. As mentioned above, the limit of RFT applicability is well described by the condition $\kappa (kl)\approx 0.5\epsilon^{-1}$. For large amplitudes such that $\kappa (kl) > 0.5\epsilon^{-1}$ the RFT can significantly overestimate the net swimming distance, as can be seen from the results in Fig.~\ref{fig:mean_k}\emph{a} corresponding, in particular, to $kl=9.25$ ($\vartriangleright$) and $kl=14$ ($\vartriangle$). The global maximum  advancement $D/l\approx 0.115$ is achieved for $kl=9.25$ ($\vartriangleright$) and $b/l\approx 0.2$ corresponding to $p\simeq 0.82$ in accord with the results shown in Fig.~\ref{fig:mean_p}\emph{a}. The animation of the particle-based undulatory swimmer corresponding to the maximum displacement-per-stroke is provided in the supplementary material (see Movie \#1).

Note that smaller amplitude $b/l$ is required for the furthest displacement upon increasing $kl$ (\ie decreasing wavelength of the undulations). This trend is in qualitative agreement with the experimental findings (\eg see Fig.1c. in \cite{berri2009}) whereas the \emph{C. elegans} undulation waveform was modulated by interaction with the motility medium by varying concentration of a thickening agent (gelatine). However, the theoretically predicted optimal amplitudes are about four folds higher than these reported in \cite{berri2009}, \eg for $kl\approx 3.6$ and $kl\approx 12.5$, the experimentally observed amplitudes were $b/l\approx 0.25$ and $\approx 0.05$, respectively, while we found (for about the same values of $kl$, see Fig.~\ref{fig:mean_k})\emph{a} the optimal amplitudes are $b/l\approx 1$ and $0.2$. It should be noticed that gelatin solutions in \cite{berri2009} exhibited viscoelastic behavior and the present theory cannot be applied directly to analyze these experimental results.

The maximum pitching angle $\theta_{max}$ is depicted vs. the scaled amplitude $b/l$ in Fig.~\ref{fig:mean_k}\emph{b} for the same values of $kl$ as in the previous figure. The agreement with the RFT prediction is excellent for all values of $kl$ and $b/l$ and does not seem to suffer from the non-local nature of the hydrodynamic self-interaction of the curved filament. We argue that since the pitching angle is just the maximum of the integral over the angular velocity, while the traveled distance combines translation and pitching, the latter is expected to be more sensitive to the intra-filament hydrodynamic interaction.

The scaled optimal work per period, $\mathcal{W}/f_{||}c^2 l T$, is depicted vs. the dimensionless amplitude, $b/l$, in Fig.~\ref{fig:mean_k}\emph{c} for the same values of $kl$ as in the previous two figures. The work is a monotonically growing function of the amplitude, while there is a crossover to a much more sluggish growth at some value of the amplitude $b/l$ depending on $kl$, which roughly corresponds to having less than half-wave ($p\approx0.5$) in the waveform. It can be explained intuitively by the fact that for $p \lesssim0.5$ the swimming gait no longer resembles traveling wave; describing it as such is misleading even though mathematically correct. The agreement between the RFT and the particle-based computations is very good for low values of $kl$, while RFT underestimates the power at large amplitudes and this deviation increases with $kl$.
\begin{figure}[t]
\begin{center}
\begin{tabular}{cc}
\includegraphics[scale=0.85]{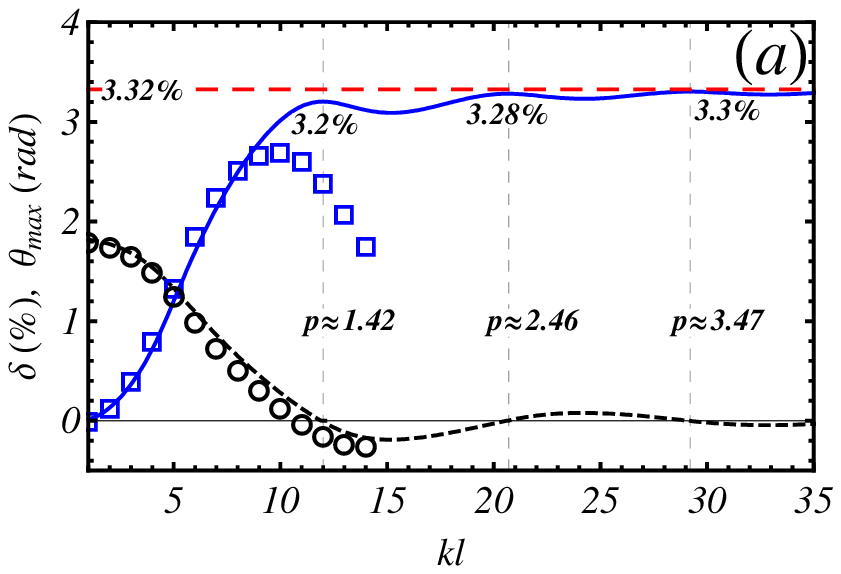} & 
\includegraphics[scale=0.85]{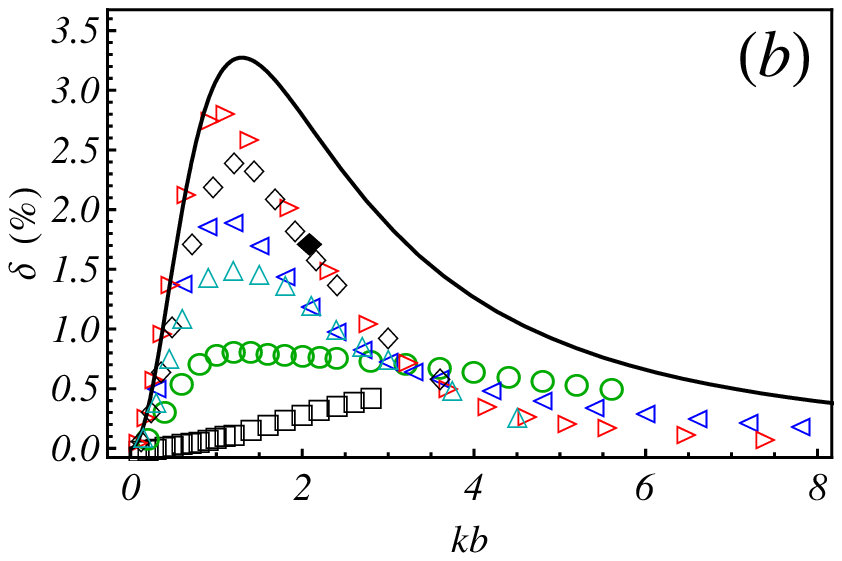}   
\end{tabular}
\end{center}
\caption{(\emph{a}) Hydrodynamic propulsion efficiency, $\delta$(\%) (blue solid line), and maximum pitching angle, $\theta_{max}$ (black dashed line) vs. $kl$ from the RFT for a finite filament with $\xi=1.515$ and amplitude-to-wavelength ratio $\kappa=kb$ being fixed at the value $1.29$ which corresponds to the RFT efficiency peak of an infinite filament with the same $\xi$ ($\delta\simeq 3.32$~\%, red dashed line); vertical (thin) dashed lines mark the location of the local optima; symbols ($\square$ , $\circ$) stand or the particle-based calculations; (\emph{b}) hydrodynamic efficiency, $\delta$, vs. $\kappa=kb$, comparison of particle-based computation results for several values of $kl$: $2$ ($\square$), $4$ ($\circ$), $6$ ($\vartriangleleft$), $9.25$ ($\vartriangleright$), $12$ ($\lozenge$) and $15$ ($\vartriangle$)  vs. the infinite filament RFT prediction (\ref{eq:eff0}) for $\xi=1.515$ (thick solid line); the filled diamond ($\blacklozenge$) stands for the efficiency of the distance-wise best performing sinusoidal waveform.
\label{fig:eff_k}}
\end{figure}

Using the results for the power invested in swimming and the displacement per stroke, we can determine the hydrodynamic propulsion efficiency $\delta$, as defined in (\ref{eq:effdef}). $\delta$ was determined via (\ref{eq:eff1}) using RFT computations for a finite filament with $\epsilon=1/30$ and $\xi=1.515$ and depicted (solid line) in Fig.~\ref{fig:eff_k}\emph{a} upon varying  $kl$ for the fixed value of $\kappa=1.29$ corresponding to the optimal infinite filament with the same $\xi$. The dashed (red) line corresponds to the optimal efficiency, $\delta\simeq 3.32$~\%, of the infinite filament based on RFT (in Eq.~\ref{eq:eff0} with $\xi=1.515$). It can be readily seen that our earlier arguments regarding the best performing (efficiency-wise) swimmer apply: for a finite filament it is advantageous to swim using short small-amplitude waves and the efficiency is growing function of $kl$ upon keeping $\kappa$ at the optimum. However, $\delta$ predicted by the RFT for the finite filament is not a monotonic function of $kl$ and there are local maxima, first of which appears at $kl\approx 12$, where $\delta\approx 3.2$\%, rather close to the optimal efficiency of the infinite filament for the same $\xi=1.515$, $\delta\simeq 3.32$~\%. The local maxima in the hydrodynamic efficiency based on RFT occur for waveforms with about half-integer number of full waves (\ie half-integer $p$) and are associated with zeros in $\theta_{max}$ as can be seen from Fig.~\ref{fig:eff_k}\emph{a}. This is in agreement with the small-$\kappa$ asymptotic predictions of $\theta_{max}$ and $\delta$ in Sec.~\ref{sec:smallkappa} even though the results shown in Fig.~\ref{fig:eff_k}\emph{a} correspond to $\kappa>0.4$, indicating that it holds for an arbitrary $\kappa$. This is also in accord with the recent work \cite{KST12} and with \cite{SL10} where the bias towards waveforms with half-integer number of full waves was suggested to result from a competition between rotational motions and bending costs. Since we have no other costs rather than hydrodynamic dissipation, it seems that the bias towards waveforms with half-integer $p$ is just due to minimal dissipation associated with pitching.

However, particle-based calculations ($\square$ in Fig.~\ref{fig:eff_k}\emph{a}) do not show the oscillations in the efficiency predicted by the RFT and there is global optimum at finite $kl$ thanks to the hydrodynamic self-interaction of the curved filament. The efficiency computed from particle-based calculations is shown in a separate Fig.~\ref{fig:eff_k}\emph{b} vs. $\kappa$ for several values of $kl$. The thick solid line stands for the estimate (\ref{eq:eff0}) for an infinite filament with $\epsilon=1/30,\:\xi=1.515$. The peak efficiency $\delta\approx2.8$~\% is achieved at $kl\approx9.25$, $b/l\approx0.12$ corresponding to $\kappa\approx1.11$ and $p\approx 1.2$, the animation corresponding to the most efficient particle-based swimming gait is provided in the supplementary material (Movie \#2). The curves for higher values of $kl \gtrsim 9.25$ yield lower values of $\delta$ at their peak in contrast to the intuitive argumentation provided in the introduction. Note also that the best swimming gait based on distance covered in a period is somewhat less efficient -- its hydrodynamic efficiency $\delta \approx 1.7$\%, and it requires similar wavelength $kl\approx 8.7$ with doubled amplitude of $b/l\approx0.24$ corresponding to $p\approx 0.8$ ($\blacklozenge$ in Fig.~\ref{fig:eff_k}\emph{b}).  On the other hand, the most efficient swimmer performs quite well in terms of swimming distance, as $D/l \simeq 0.093$ vs. $D/l\simeq 0.12$ at the optimum. Therefore, keeping the undulation amplitude $b/l$ and the wavelength $kl$ in the range $0.12\lesssim b/l\lesssim 0.24$, $8.7 \lesssim kl \lesssim 9.2$, respectively, would yield good swimming performance both speed- and power-wise.

In comparison to a considerable rotation ($\theta_{max}\approx 53^\circ$) associated with the  distance-wise optimal swimming gait, when hydrodynamic efficiency is optimized the pitching is quite small with $\theta_{max} \approx 9.5^\circ$ (see Fig.~\ref{fig:mean_p}\emph{b}). To the best of our knowledge, the displacements featured by microorganisms do not exhibit visibly apparent pitching; this may be however due to factors such as more complicated non-sinusoidal waveforms thanks to complex mechanosensory mechanisms including proprioception (\ie, the sense of the body's curvature and the strength employed in movements) \cite{li2006}.

\subsection{Nematode C. elegans\label{sec:celegans}}
\begin{figure}[t]
\begin{center}
\begin{tabular}{cc}
\includegraphics[scale=0.4]{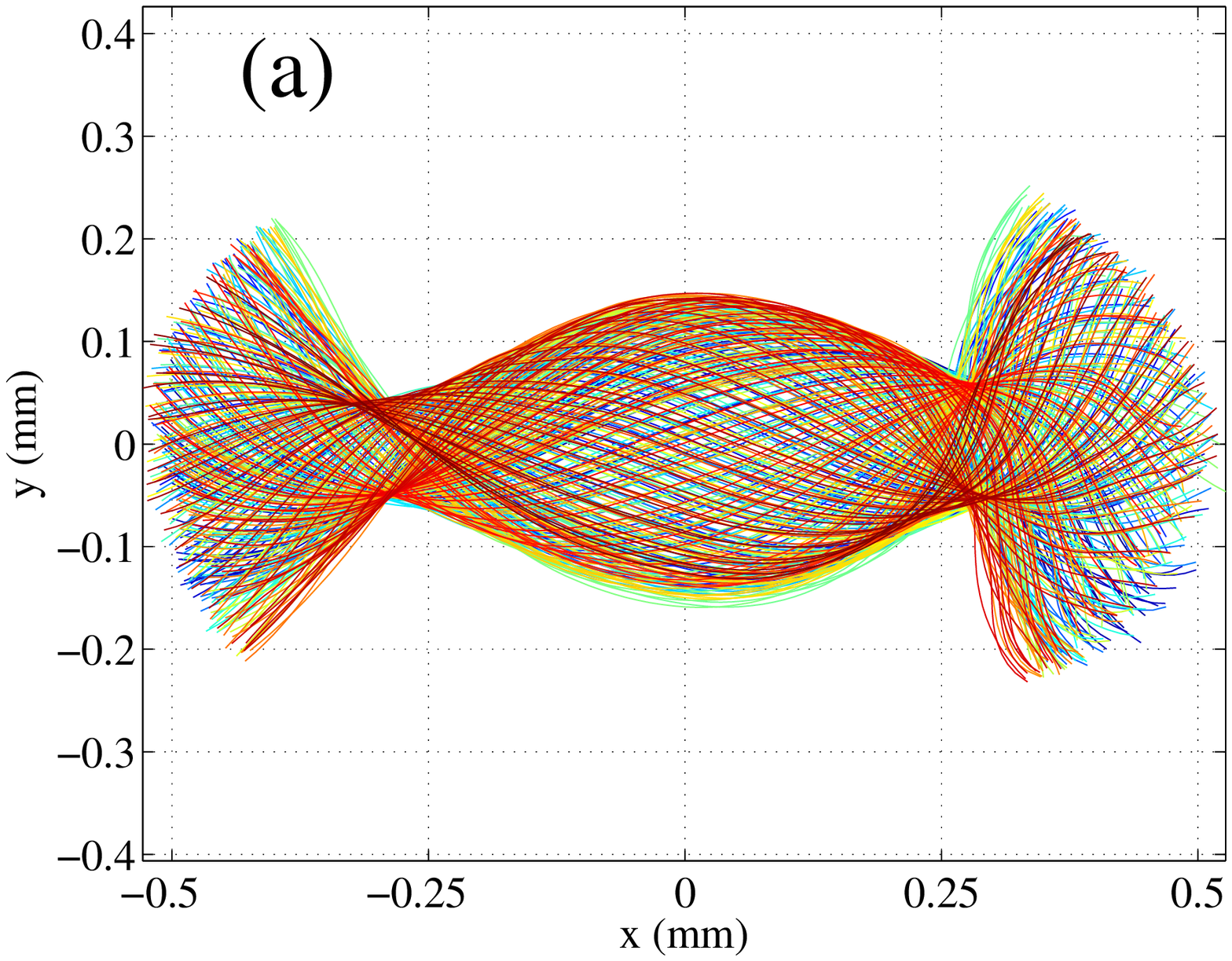} & %
\includegraphics[scale=0.4]{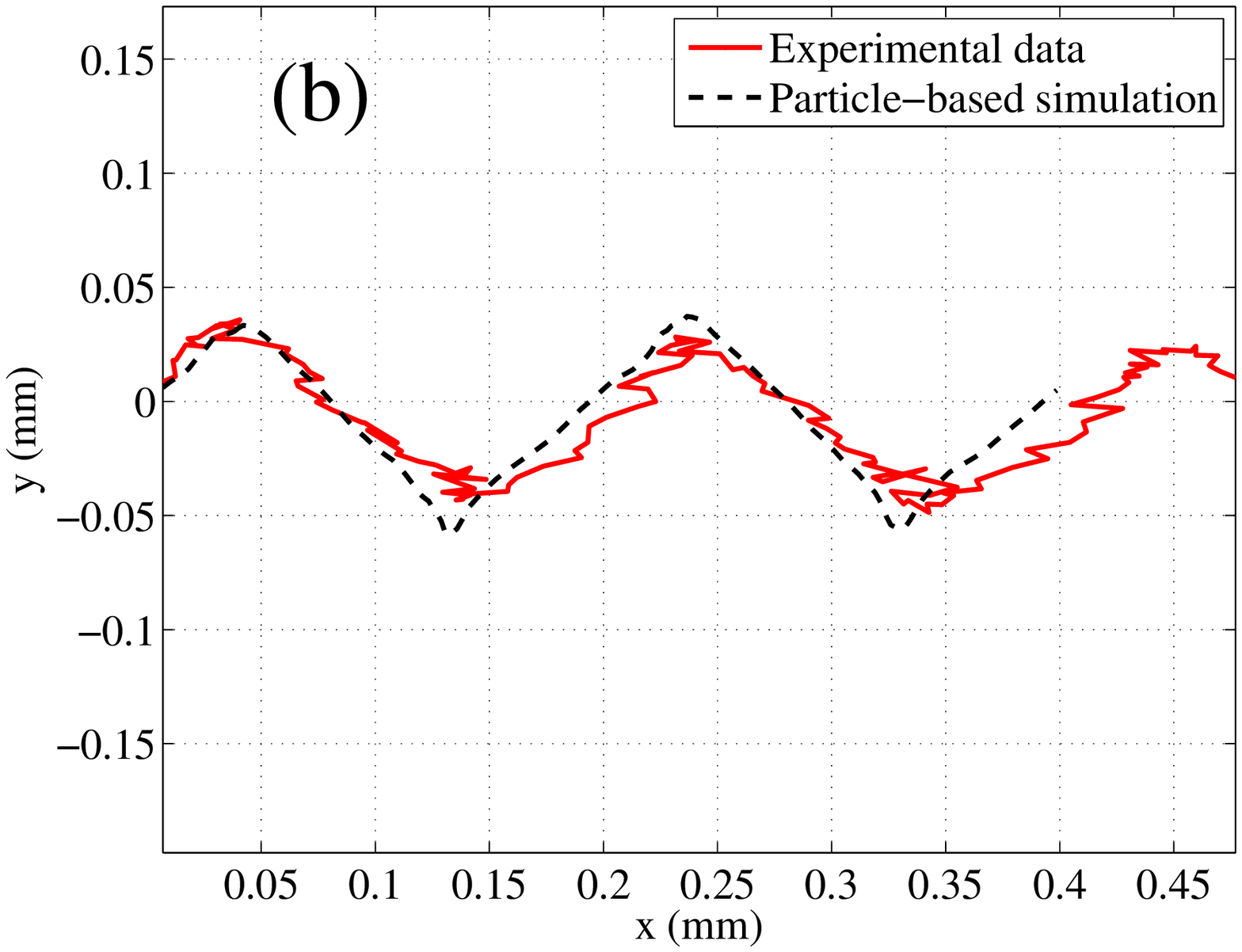}
\end{tabular}
\end{center}
\caption{(\emph{a}) The snapshots of the nematode waveforms in the co-rotated and co-moving frame of reference as tracked in the experiment in \cite{celegans10}; the worm is propelled head to the right (\emph{b}) The trajectory of the geometric center of the nematode: tracking experiment (solid line) and particle-based simulation (dashed line). The worm in this experiment was 1.2~mm long, and it progressed 0.2~mm per period of undulation, yielding $D/l\approx 0.17$. \label{fig:exper}}
\end{figure}

The particle-based algorithm was applied to model propulsion of the nematode \emph{C. elegans}. We use a filament built from $N=12$ spheres to mimic the nematode slenderness of $\epsilon=0.083$ (typical length of $1$~mm and width of $0.08$~mm). The swimming gait adopted in computation was extracted from videos shot with a high-speed camera via the use of custom-written image processing algorithm \cite{celegans10}. The snapshots of the nematode waveforms (in the co-rotating and co-moving frame) are shown in Fig.~\ref{fig:exper}\emph{a}. It can be readily seen that the nematode waveform is not a sinusoidal wave and that the undulation amplitude varies along the body length so that the head and the tail's amplitudes are larger than that in the middle portion of the body. The animation of the particle-based nematode swimming is provided in the supplementary material (Movie \#3).

The comparison between the experimentally probed trajectory and the numerically calculated path that uses the tabulated undulation gait extracted from experiments (in Fig.~\ref{fig:exper}\emph{a}) is shown in Fig.~\ref{fig:exper}\emph{b}. The close agreement between the experimental and the numerical results (with no adjustable parameters) justifies the use of low Reynolds hydrodynamics in \emph{C. elegans} locomotion study whereas typically $Re\sim 1$ in low viscosity aqueous medium. The typical parameters for \textit{C. elegans} propulsion are $b/l\approx 0.12$, $kl\approx7.9$ and $D/l\approx0.17$ as reported in \cite{celegans10}. Using the particle-based scheme, the power invested in the nematode swimming per period was determined giving an unexpectedly high hydrodynamic efficiency of $\sim$ 8.8\%\footnote{Note that the optimal swimming efficiency corresponding to Lighthill's sawtooth traveling wave propagating along an infinite filament is $\delta=8.58$\% \cite{lighthill75}} and low pitching with $\theta_{max}\approx 8.3^\circ$, comparable to the most efficient sine swimmer ($\theta_{max}\approx 9.5^\circ$). The power invested in swimming was calculated in the same way as before, whereas the longitudinal drag coefficient $f_{||}\approx 6.25\mu$ was determined by fitting results of particle-based computations to RFT for a filament with aspect ratio $\epsilon\simeq 1/12$ propelled using sine waveform. Even though the nematode does not use a simple sine wave, the parameters of the sine waveform optimized to the furthest advancement per stroke are similar to the values employed by the nematode (see the comparison in Fig.~\ref{fig:organisms}\emph{a}). However, the shape of the waveform exploited by the worm allows a superior locomotion (in terms of both the displacement per stroke and hydrodynamic efficiency) compared with the sine waveform optimized for the furthest displacement showing $D/l\approx 0.12$ and $\delta \simeq 1.7$~\%.

Note also that using the typical parameters for \emph{C. elegans} propulsion in low viscosity medium ($kl\approx 7.9$ and $b/l \approx 0.12$, see \cite{celegans10}) and the aspect ratio of $\epsilon\approx 1/12$ gives $\kappa (kl)\approx 7.5$, while $1/(2\epsilon)\approx 6$. Therefore, the estimate indicates that the RFT, widely used to model \emph{C. elegans} swimming, may not be accurate from a hydrodynamic point of view, and models accounting for non-local hydrodynamic interaction, such as particle-based algorithm, should be invoked.

\subsection{Common undulatory microswimmers}
\begin{figure}[t]
\begin{center}
\begin{tabular}{cc}
\includegraphics[scale=0.9]{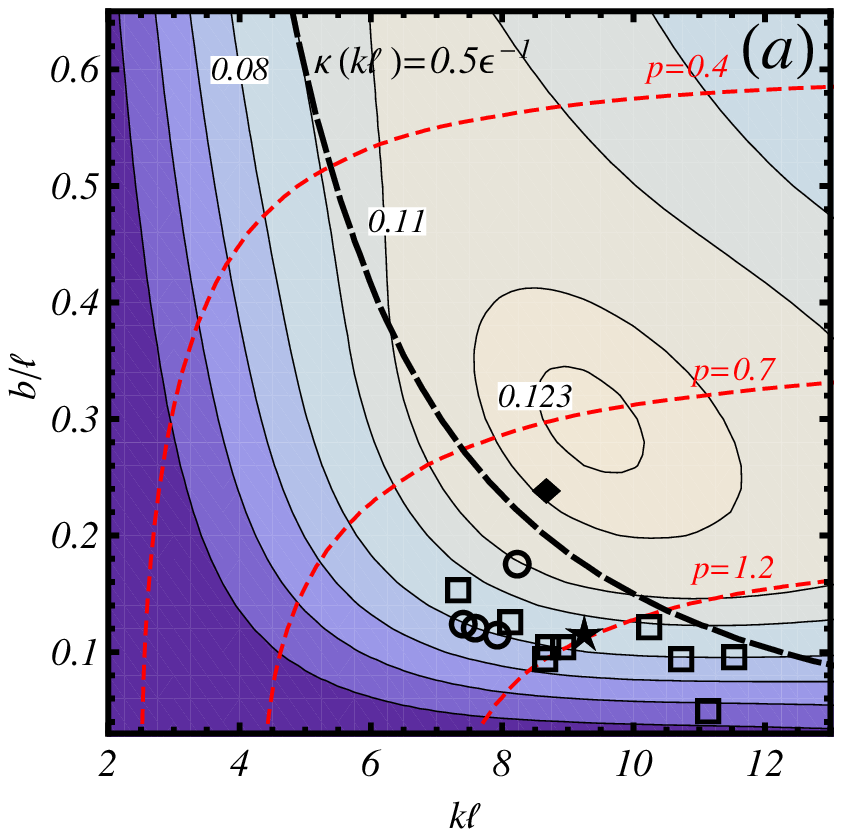} & 
\includegraphics[scale=0.9]{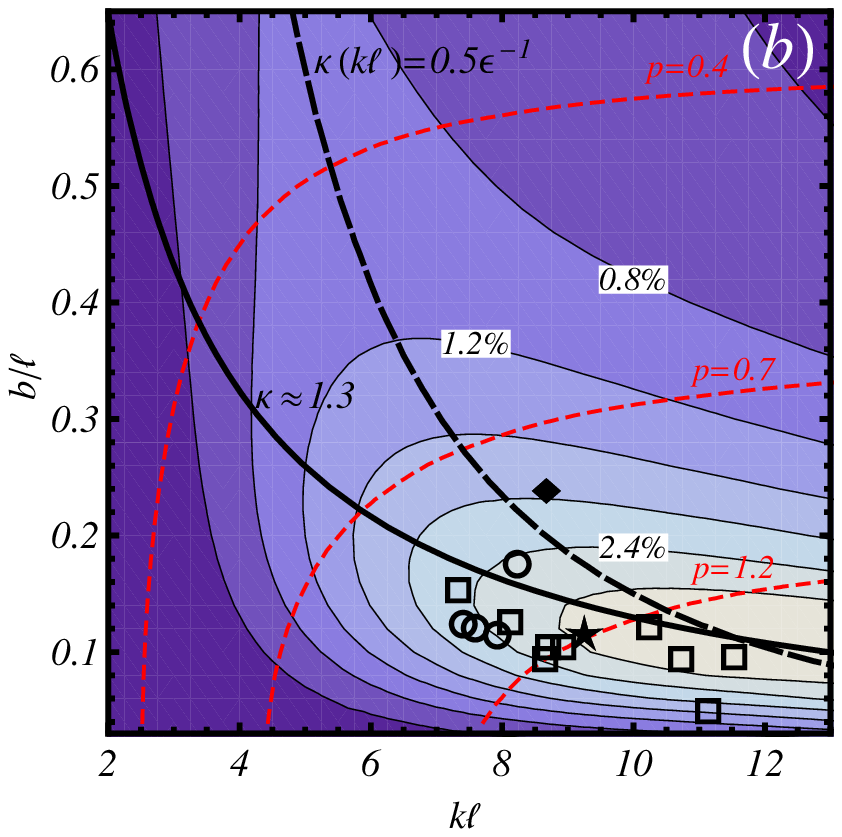} 
\end{tabular}
\end{center}
\caption{Color contour plots based on prediction of local RFT for a filament of the aspect ratio $\epsilon=1/30$ in plane of parameters $(b/l,\, kl)$. The corresponding sections of fixed mean $p$ are shown as thin dashed lines (red, short dashes). Labels (on white background) stand for some representative values along the contour lines. The thick dashed line (black, long dashes) stands for the boundary of approximate validity of the RFT, \ie $\kappa (kl)=0.5 \epsilon^{-1}$. The squares ($\square$) stands for the available data for sperm cells (\emph{P. miliaris} \cite{gray1955}; \emph{bos}, \emph{chaetopterus}, \emph{ciona}, \emph{colobopocentotrus}, \emph{lytechinus}, \emph{psammechimus} \cite{brennen1977}; \emph{ostrea}, \emph{ovis} \cite{DENEHY}, the circles ($\circ$) stand for the data for nematodes (\emph{C. elegans} \cite{sznitman2010}; \emph{Haemonchus contortus}, \emph{Turbatrix aceti}, \emph{Pamagrellus silusia} \cite{gray1964}), filled symbols $\blacklozenge$ and $\bigstar$ denote the best, distance-wise and efficiency-wise gaits, respectively, determined in particle-based computations; (a) scaled distance per period of undulation, $D/l$; (b) hydrodynamic efficiency $\delta$; the thick solid line stands for the location of the optimal $\delta$ for an infinite filament based on Eq.~\ref{eq:eff0}, \ie $\kappa\approx 1.3$.     \label{fig:contours}}
\end{figure}

Provided a favorable agreement between the particle-based numerical results and the prediction of the RFT, and since the later approach is not as time-consuming as the particle-based simulation algorithm, we applied the finite filament RFT to calculate the scaled displacement per cycle, $D/l$, and hydrodynamic efficiency $\delta$ for a filament with aspect ratio $\epsilon=1/30$ in a wide range of parameters $(b/l,kl)$ and depicted the results in  Figs.~\ref{fig:contours}\emph{a},\emph{b} as a color contour plots. The dashed (red, short dashes) lines in Figs.~\ref{fig:contours}\emph{a,b} are the cross-sections corresponding to fixed value of $p$ (as in Fig.~\ref{fig:mean_p}). The thick (black, long dashes) line stands for the approximate boundary of the RFT validity as discussed above, $\kappa (kl)=0.5\epsilon^{-1}$. Above this curve the prediction of the RFT may considerably overestimate the swimming distance and hydrodynamic efficiency and more accurate estimates taking into account non-local hydrodynamic interaction should be invoked such as, for example, particle-based computations.

As can be seen in Fig.~\ref{fig:contours}\emph{a} the RFT predicts the maximum swimming distance  $D/l \approx 0.124$ at $kl\simeq 9.2$, $b/l \simeq 0.3$ corresponding to $p \approx 0.7$, while according to particle-based computations the optimum ($D/l\approx 0.117$) is at $b/l\approx 0.24$, $kl\approx 8.7$ corresponding to $p\approx0.8$ (black diamond $\blacklozenge$ in Figs.~\ref{fig:contours}\emph{a,b}). As discussed before, the global optimum in the hydrodynamic efficiency based on RFT ($\delta \simeq 3.32$~\%) is achieved for an infinite filament, \ie at $kl \rightarrow \infty$, $b/l \rightarrow 0$ at $kb\simeq 1.3$, however, a local optimum ($\delta \approx 3.2$~\%) is achieved at $kl \approx 12$ and $b/l\approx 0.11$ (see Fig.~\ref{fig:eff_k}\emph{a}). Note that the thick solid line corresponding to $\kappa=1.3$ which maximizes $\delta$ in case of infinite filament, crosses the (white) region of close-to-optimal propulsion efficiency of a finite filament in Fig.~\ref{fig:contours}\emph{b}. For comparison, the most efficient swimming gait determined from particle-based computations ($kl \approx 9.25$ and $b/l \approx 0.12$ corresponding to a waveform with $p \approx 1.2$ complete waves) maximizing the hydrodynamic efficiency, $\delta\simeq 2.8$~\%, is marked by a star symbol ($\bigstar$) in Figs.~\ref{fig:contours}\emph{a,b}.

\begin{figure}[t]
\begin{center}
\begin{tabular}{cc}
\includegraphics[scale=0.9]{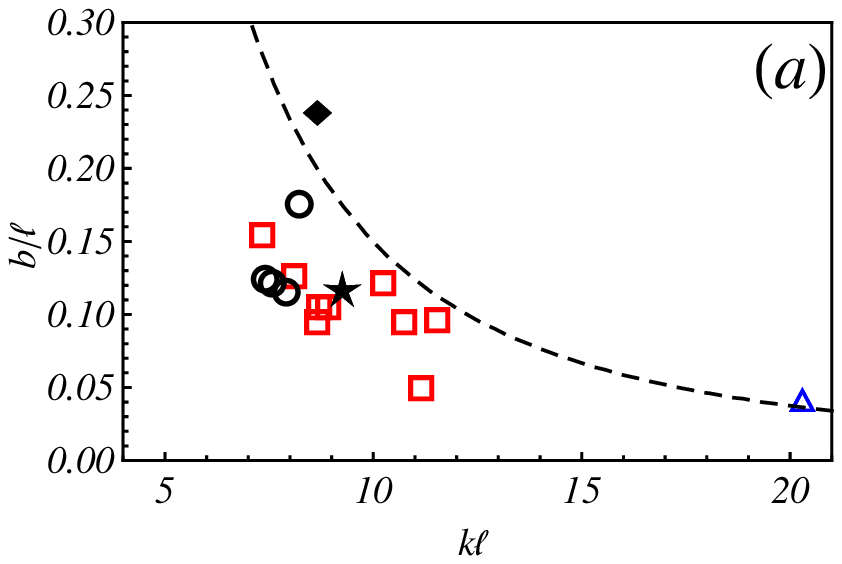} & 
\includegraphics[scale=0.9]{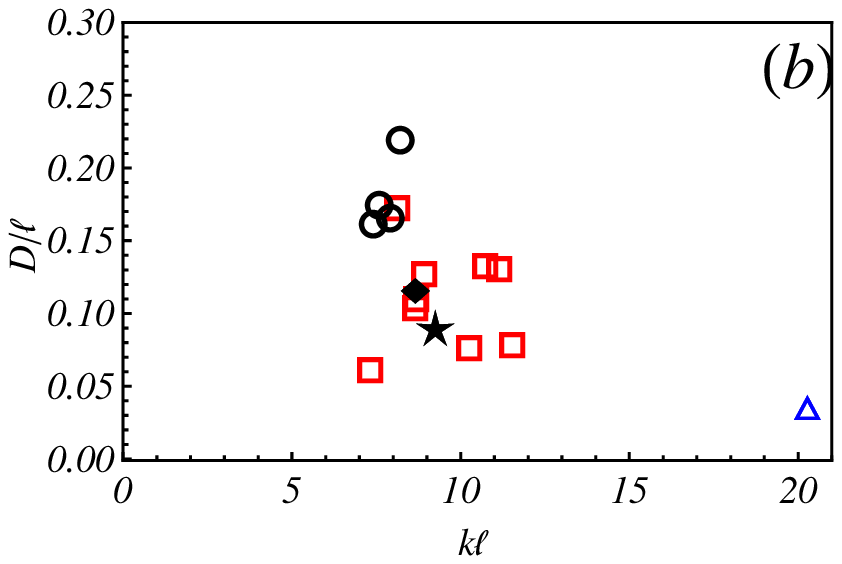} 
\end{tabular}
\end{center}
\caption{The comparison of actuation parameters and performance of undulating microorganisms (empty symbols) vs. a best-performing distance- ($\blacklozenge$) and efficiency-wise ($\bigstar$) filament with aspect ratio $\epsilon=1/30$ propelled by a sinusoidal waveform determined from particle-based computations. The red squares ($\square$) correspond to sperm cells (\emph{P. miliaris} \cite{gray1955}; \emph{bos}, \emph{chaetopterus}, \emph{ciona}, \emph{colobopocentotrus}, \emph{lytechinus}, \emph{psammechimus} \cite{brennen1977}; \emph{ostrea}, \emph{ovis} \cite{DENEHY}, the black circles ($\circ$) stand for nematodes (\emph{C. elegans} \cite{sznitman2010}; \emph{Haemonchus contortus}, \emph{Turbatrix aceti}, \emph{Pamagrellus silusia} \cite{gray1964}) and the blue upper triangle ($\vartriangle$) is an eukaryote flagellate (\emph{Ochromonos malhamensis} \cite{brennen1977}); (a) dimensionless amplitude of undulations $b/l$ vs. wave number $kl$; the dashed line corresponds to the approximate boundary of RFT validity, \ie $\kappa (kl)=0.5 \epsilon^{-1}$ for $\epsilon=1/30$; (b) scaled swimming distance per period, $D/l$ vs. scaled wave number $kl$.  \label{fig:organisms}}
\end{figure}

The empty squares ($\square$) in both Figs.~\ref{fig:contours}\emph{a,b} correspond to sperm cells  \cite{brennen1977,gray1955,DENEHY,Friedrich}, empty circles ($\circ$) to nematodes \cite{gray1964,sznitman2010}, all swimming using periodic undulations. These data together with an eukaryote flagellate \cite{brennen1977} ($\vartriangle$) and the corresponding values of displacement per period, $D/l$, are shown in two separate Figs.~\ref{fig:organisms}\emph{a},\emph{b}.  It is evident that even though the microorganisms do not exploit a sine waveform for propulsion (\eg see Fig.~\ref{fig:exper}\emph{a} that illustrates the swimming gait of \emph{C. elegans}), the typical amplitudes and wavelengths they exploit are quite close to the best distance- and, in particular, efficiency-wise sine-waveform swimming gait determined from particle-based computations ($\blacklozenge$ and $\bigstar$, respectively). Similarly to Fig.~\ref{fig:contours}\emph{a} the dashed line in Fig.~\ref{fig:organisms}\emph{a} marks the approximate boundary of validity of the RFT for a mode sine swimmer with $\epsilon=1/30\simeq 0.033$. The nematodes (\eg \emph{Haemonchus contortus}, \emph{Turbatrix aceti}, \emph{Pamagrellus silusia}) in \cite{gray1964} have $\epsilon$ in the range $0.03$--$0.04$ (except for \emph{C. elegans} with $\epsilon \approx 0.083$ \cite{celegans10}), while sperm cells are typically more slender, \eg ram and oyster sperm cells \cite{DENEHY} having $\epsilon\approx 0.008$ and $0.005$, respectively. Therefore, the use of RFT for most nematodes and sperm cells is probably justified, while for less slender swimmers, such as \emph{C. elegans}, it may not produce accurate results as discussed above in Sec.~\ref{sec:celegans}.

\begin{figure}[t]
\begin{center}
\begin{tabular}{cc}
\includegraphics[scale=0.9]{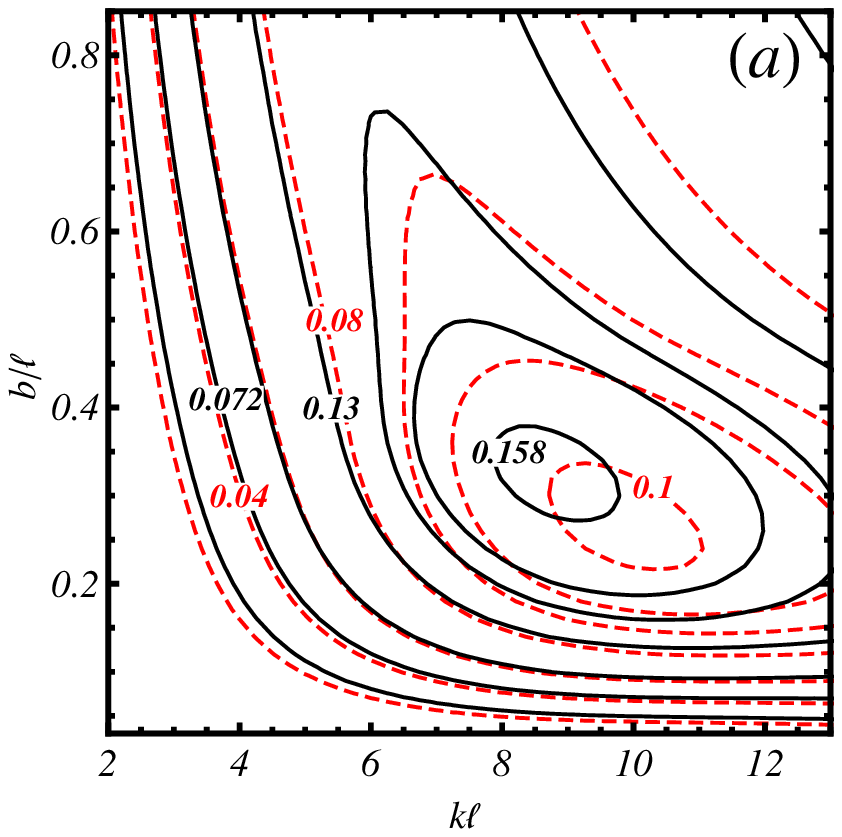} & 
\hskip 0.2cm
\includegraphics[scale=0.9]{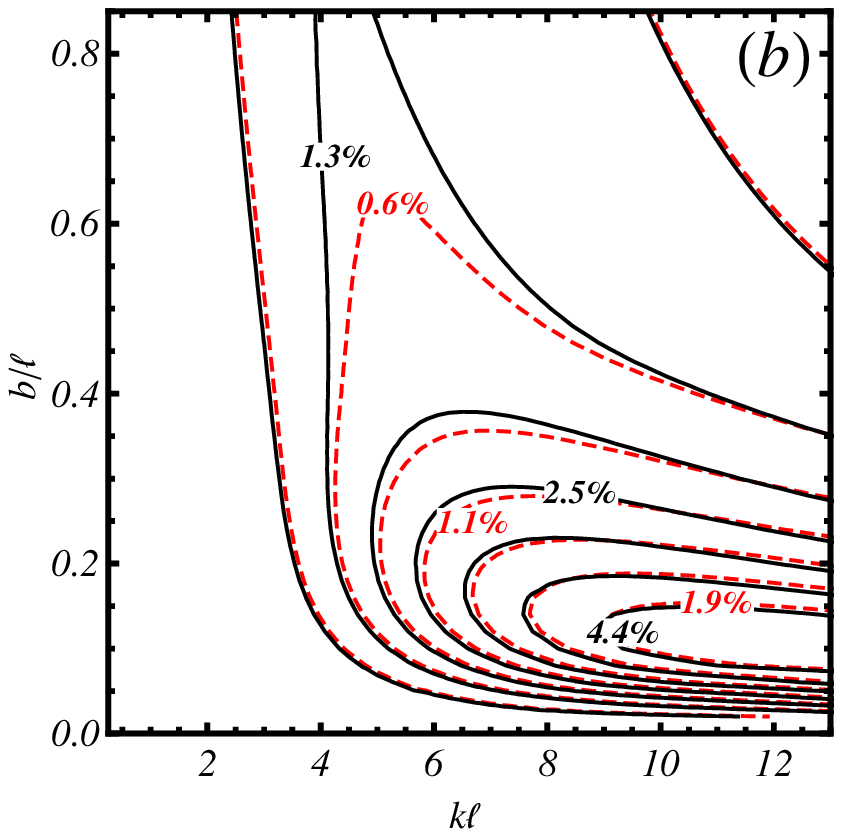} 
\end{tabular}
\end{center}
\caption{The effect of filament slenderness on the swimming performance as predicted by the RFT in plane of parameters $kl$ and $b/l$; dashed (red) curves stand for the filament with aspect ratio $\epsilon=1/12$ ($\xi=1.4$), solid (black) lines correspond to a filament with $\epsilon=1/800$ ($\xi=1.7$). Contour labels depict some representative values along selected isolines (\emph{a}) contour plots of the swimming distance per stroke, $D/l$; (\emph{b}) contour plots of hydrodynamic propulsion efficiency $\delta$.
   \label{fig:contours1}}
\end{figure}
Varying the slenderness $\epsilon$ of the filament does not change much the topography of the surfaces in Figs.~\ref{fig:contours}\emph{a},\emph{b} as can be seen in Figs.~\ref{fig:contours1}\emph{a},\emph{b} where we plot the isolines predicted by the RFT corresponding to the swimming distance per period $D/l$ and the efficiency $\delta$ for two quite different values of slenderness, $\epsilon\sim1/12$ and $\epsilon\sim 1/800$. This weak dependence is to be expected, since in the RFT the slenderness only enters through $\xi=f_\perp/f_{||}$ which is a weak (logarithmic) function of $\epsilon$, (the value of $\xi$ increases from $\sim 1.4$ to $\sim 1.7$ as $\epsilon$ decreases from $1/12$ to $1/800$ -- by over 60 folds).

It can be readily seen that the variance in $\xi$ has only a minor effect on the location of the optima for both $D/l$ and $\delta$. The maximum distance $D/l\approx0.101$ is achieved at $kl\approx9.75$ and $b/l\approx0.28$ for $\xi=1.4$ while for $\xi=1.7$ the maximum $D/l\approx 0.159$ is achieved at $kl\approx8.75$, $b/l\approx0.32$. The peak in propulsion efficiency (using RFT) is achieved at the maximal available $kl$, namely at $kl=12.5$ (higher $kl$ will yield slightly higher efficiency) for $b/l\approx0.1$ and it is $\delta\sim2.2$\% and $\sim4.9$\% for $\xi=1.4$ and $1.7$, respectively.

\section{Concluding remarks}
In this paper we studied low-Reynolds-number locomotion of finite undulating filament of length $l$ and width $2a$ propelled by a propagating sinusoidal wave with amplitude $b$ and wavenumber $k$ using an approximate Resistive Force Theory (RFT), assuming a local nature of hydrodynamic interaction between the filament and the surrounding liquid, and more accurate particle-based numerical computations taking into account the intra-filament hydrodynamic interaction.  Based on the results, the approximate limit of RFT applicability was found as $\kappa (kl) \lesssim 1/(2\epsilon)$, where $\kappa=kb$ is dimensionless amplitude-to-wavelength ratio and $\epsilon=2a/l$ is filament's aspect ratio. For $\kappa (kl) > 1/(2\epsilon)$ the predictions of the RFT may significantly deviate from the results of particle-based computations.

We showed that there is an optimal combination of the dimensionless undulation amplitude $b/l$ and wavenumber $kl$, namely $b/l\simeq 0.24$, $kl\simeq 8.7$ yielding the furthest advancement per period of undulation, $D/l\simeq 0.12$. This propulsion gait is characterized by a waveform with $p\simeq$0.8 complete waves per filament length, considerable pitching (maximum tilt angle  $\theta_{max}\sim53^\circ$) and hydrodynamic efficiency  $\delta\simeq1.7$\%. Reducing the undulation amplitude two folds, to  $b/l\simeq 0.12$ together with $kl\simeq 9.2$ yields the most efficient propulsion with $\delta\simeq2.8$\%. The latter swimming gait is characterized by $p\simeq 1.2$ complete waves per filament length, low pitching with $\theta_{max} \sim 9^\circ$ and advancement per period $D/l\simeq 0.093$.

Comparison to the experimental results for \emph{C. elegans} reveals that even though the swimming characteristics in terms of amplitude and wavelength are quite similar to the best performing (distance- and efficiency-wise) sinusoidal swimmer's gait, the nematode is superior to the sinusoidal swimmer in terms of both the swimming distance per stroke ($D/l\simeq 0.17$) and hydrodynamic efficiency ($\delta\approx 8.8$\%) as estimated from particle-based computations exploiting the nematode swimming gait extracted from experiments. This indicates the importance of the amplitude modulation in the waveform adopted by \emph{C. elegans}, deviating considerably from a simple sine wave. Comparison to available data for other undulatory micro-swimmers including various sperm cells and nematodes, shows that most of them operate in a narrow range of wavelengths $7.5\lesssim kl \lesssim 11.5$, whereas the best performing sinusoidal swimmer ($kl\simeq 9.25$) lies well within this range. The typical amplitude, $b/l$, of many undulatory biological swimmers is within the range $0.08$--$0.16$, with the most efficient sine swimmer ($b/l\simeq0.12$) again lying well inside this range, while the fastest sine swimmer requires a larger amplitude of $b/l\simeq 0.24$. Moreover, with reference to \emph{C. elegans}, most biological swimmers overperform the fastest model sine swimmer in terms of swimming distance covered per stroke period, further emphasizing the importance of the geometric waveform optimization. Based on the approximate limit of RFT applicability derived for the model sine swimmer, the swimming gaits of most undulatory biological swimmers reported in the literature are likely to be adequately described by the RFT. Modeling of relatively short swimmers, such as \emph{C. elegans}, may necessitate the use of more rigorous hydrodynamic models accounting for non-local nature of hydrodynamic interaction between different parts of the filament.

\ack
This work was partially supported by Japan Technion Society Research Fund (AML) and and the US-Israel Binational Science Foundation via BSF grant \#2011323 (JS).

\section*{Appendix A: Useful identities}

Denoting $\langle \dots \rangle={1\over\lambda}\int_0^\lambda(\ldots)ds$,  for $\gamma={\rd\zeta\over \rd s}=(\hat{\bbs}\cdot\hat{\bx})^{-1}=\sqrt{1+\kappa^2\cos^2(k s-\it{\Omega} t)}$ we have:
$$\langle\gamma\rangle={2\over\pi}\mathrm{E}(-\kappa^2)\:,$$
$$\langle{1\over\gamma}\rangle={2\over\pi}\mathrm{K}(-\kappa^2)\:,$$
$$\langle\gamma^2\rangle=1+\kappa^2/2\:,$$
$$\langle{1\over\gamma^2}\rangle={1\over2\sqrt{1+\kappa^2}}\:.$$


\section*{Appendix B: Particle-based computation scheme}

The general solution for the velocity and the pressure field around a collection of $N$ spherical particles of radii $\mathrm{a}_i$, can be written as
\begin{equation}
\mathbf{v}=\sum_{i=1}^{N}{\bv_i}\:,\quad p=\sum_{i=1}^{N} P_i
\end{equation}
where the solution for the velocity $\bv_i$ outside a single $i$th sphere has the form of Lamb's general solution of Stokes equations in terms of solid spherical
harmonics \cite{KK91},
\begin{eqnarray}
\bv_i&=&\bv_i'+\frac{1}{2\mu} \bbr_i P_i=\sum_{n=1}^{\infty}{\nabla \times \left(\bbr_i \chi^{i}_{-\left(n+1\right)}\right)+\nabla \Phi^i_{-\left(n+1\right)}}- \nonumber \\
&& \frac{\left(n-2\right)}{\mu 2n\left(2n-1\right)}r_i^2 \nabla p^i_{-\left(n+1\right)}+\frac{\left(n+1\right)}{\mu n \left(2n-1\right)}\bbr_i
p^i_{-\left(n+1\right)}
\end{eqnarray}
Here $\bbr_i$ is the radius vector with origin at the center of the $i$th sphere, $r_i=|\bbr_i|$, $p^i_{-(n+1)}$ is a linear combination of solid spherical harmonics of order $-(n+1)$ with the origin at the center of the $i$th sphere, satisfying the Laplace equation for the pressure field $\nabla^2 P_i=0$, while $\chi^i_{-(n+1)}$, $\Phi^i_{-(n+1)}$ each are combinations of solid harmonics, arising from the solution of the associated homogeneous equations $\nabla\cdot \bv_i=0$ and $\nabla^2 \bv_i'=0$:
\beq
\left\{\Phi^{i}_{-\left(n+1\right)},\frac{1}{\mu}p^i_{-\left(n+1\right)}, \chi^i_{-\left(n+1\right)} \right\}=\sum_{m=-n}^n \left\{ a_{mn}^i, b_{mn}^i, c_{mn}^i\right\}\: u_{mn}^{i-}\, ,\nonumber \\
\eeq
with $u^{i-}_{mn}$ being decaying solid spherical harmonics centered at the origin of the $i$th sphere
\beq
u_{mn}^{i-}=\frac{1}{r_i^{n+1}}P_n^m\left(\cos{\theta_i}\right)\re^{\ri m \phi_i},\label{eq:harmonics}
\eeq
where $P_n^m$ is the associated Legendre function. For $n=1$ the solution $\{\Phi^i_{-2},\,\frac{1}{\mu} p^i_{-2},\,\chi^i_{-2}\}$ corresponds, respectively, to a stresslet, stokelet and rotlet centered at the $i$th sphere \cite{KK91}.

The no-slip boundary conditions, $\bv=\bu_i$, where $\bu_i$ is the local velocity of the surface of $i$th particle, can be used to determine the unknown coefficients $a_{mn}^i$, $b_{mn}^i$ and $c_{mn}^i$. An elegant way of computing the coefficients was proposed in \cite{filippov00}. The boundary conditions are first transformed to the Lamb's form by applying operators $\bbr_i\cdot$, $-r_i\nabla\cdot$ and $\bbr_i\cdot\nabla\times$ to both sides of the no-slip boundary condition and then the direct origin-to-origin transformation of spherical harmonics centered at different spheres is applied, yielding an infinite system of linear equations for the coefficients,
\begin{eqnarray}
\label{fil1}
&-&(n+1)a_{mn}^i+\frac{(n+1)}{2\:(2n-1)}b_{mn}^i \nonumber \\
&& +\mathtt{a}_i^{2n+1}\sum_{j=1}^{N}{\sum_{l=1}^{\infty}{\sum_{k=-l}^l{\left(D_{klmn}^{ij}a_{kl}^j+E_{klmn}^{ij}b_{kl}^j+F_{klmn}^{ij}c_{kl}^j\right)}}}=
\mathtt{a}_i^{n+1} X_{mn}^i,
\end{eqnarray}
\begin{eqnarray}
\label{fil2}
&&{1\over \mathtt{a}_i^2}(n + 1)(n+2)a_{mn}^i-\frac{n(n+1)}{2\left(2n-1\right)}b_{mn}^i \nonumber \\
&&\quad  +\sum_{j=1}^{N}{\sum_{l=1}^{\infty}{\sum_{k=-l}^l{\left(G_{klmn}^{ij}a_{kl}^j+H_{klmn}^{ij}b_{kl}^j+L_{klmn}^{ij}c_{kl}^j\right)}}}=\mathtt{a}_i^n Y_{mn}^i,
\end{eqnarray}
\begin{equation}
\label{fil3}
n\left(n+1\right)c_{mn}^i+\mathtt{a}_i^{2n+2} \sum_{j=1}^{N}{\sum_{l=1}^{\infty}{\sum_{k=-l}^l{\left(M_{klmn}^{ij}b_{kl}^j+
N_{klmn}^{ij}c_{kl}^j\right)}}}=\mathtt{a}_i^{n+1} Z_{mn}^i\:.
\end{equation}
The coefficients $D_{mnkl}^{ij},\: E_{mnkl}^{ij},\: F_{mnkl}^{ij},\:K_{mnkl}^{ij},\: L_{mnkl}^{ij},\:M_{mnkl}^{ij}$ and $N_{mnkl}^{ij}$ are given in
the appendix of \cite{filippov00} in terms of
the transformation coefficient $C^{ij}_{klmn}$:
\[
C^{ij}_{klmn}=(-1)^{m+n}\frac{(l+n-k+m)!}{(l-k)!(m+n)!} u^{j-}_{(k-m)(l+n)}(R_{ij},\theta_{ij},\varphi_{ij})\:,
\]
where $R_{ij},\theta_{ij},\varphi_{ij}$ are the spherical coordinates of vector $\bR_{ij}$ connecting the centers of $j$th and $i$th spheres, $u^{j-}_{(k-m)(l+n)}$ is the decaying solid spherical harmonics defined in (\ref{eq:harmonics}). According to definition of spherical harmonics the coefficients $C_{klmn}$ are assumed zero if $|k|>l$ or if $|m|>n$.

$X_{mn}^i$, $Y_{mn}^i$ and $Z_{mn}^i$ are the coefficients in the expansions in surface harmonics of ${\bbr_i\over r_i} \cdot \bu_i$, $-r_i\nabla\cdot \bu_i$ and $\bbr_i\cdot\nabla\times \bu_i$. When the particle surface velocity corresponds to the rigid body motion, $\bu_i=\bV_i+\te{\omega}_i\times\bbr_i$, the right hand side of (\ref{fil1}-\ref{fil3}) can be written as \cite{filippov00}:
\begin{eqnarray}
\label{X11}
X_{1n}^i&=&\frac{1}{2}\left(V_{ix}^0-\ri V_{iy}^0\right)\delta_{n}^1\\
\label{X01}
X_{0n}^i&=&V_{iz}^0\delta_{n}^1\\
\label{X_11}
X_{-1n}^i&=&-\left(V_{ix}^0+\ri V_{iy}^0\right)\delta_{n}^1\\
\label{Ymn}
Y_{mn}^i&=&0\\
\label{Z11}
Z_{1n}^i&=&\left(\omega_{ix}^0-\ri \omega_{iy}^0\right)\delta_{n}^1\\
\label{Z01}
Z_{0n}^i&=&2\omega_{iz}^0\delta_{n}^1\\
\label{Z_11}
Z_{-1n}^i&=&-2\left(\omega_{ix}^0+\ri \omega_{iy}^0\right)\delta_{n}^1
\end{eqnarray}
with $\{\bV_i, \mbox{\boldmath$\omega$}_i\}$ being the translation and rotation velocities of $i$th sphere, respectively and $\delta^k_n$ being the Kronecker's delta.

The viscous drag force $\bF_i$ exerted on sphere $i$ and hydrodynamic torque $\bT_i$ about its center can be expressed in terms of the expansion coefficients for $n=1$,
\begin{eqnarray}
\label{filforce}
\bF_i=-4\pi\mu
\left[\left(b_{11}^i-\frac{1}{2}b_{-11}^i\right)\hat{\bx}+\ri\left(b_{11}^i+\frac{1}{2}b_{-11}^i\right)\hat{\by}+b_{01}^i
\hat{\bz}\right]\\
\label{filtorque}
\bT_i=-8\pi\mu
\left[\left(c_{11}^i-\frac{1}{2}c_{-11}^i\right)\hat{\bx}+\ri\left(c_{11}^i+\frac{1}{2}c_{-11}^i\right)\hat{\by}+c_{01}^i
\hat{\bz}\right]
\end{eqnarray}

Thus when velocities of the spheres are prescribed the forces and torques exerted on any sphere can be found by solving $3\:N\times L\times(L+2)$ equations for the expansion coefficients $\{a_{mn}^i, b_{mn}^i, c_{mn}^i\}$, obtained by truncating the system (\ref{fil1}-\ref{fil3}) after $l=L$ terms and solving it together with (\ref{filforce}-\ref{filtorque}). Alternatively, forces and torques can be prescribed and velocities are computed or a mixed problem can be formulated when some velocities and forces/torques are prescribed.

\end{document}